\expandafter\edef\csname hypers@fe\endcsname{\catcode
                                             `\noexpand @=\the\catcode`\@}%
\catcode`\@=11
%
%
\ifx\hyperd@ne\hyper@ndefined
 \global\let\hyperd@ne=\relax
\else
 \errhelp{hyperbasics.tex needs to be included only once outside
          of any {...} or \begingroup...\endgroup. You have tried to
          include it more than once. If the previous include was indeed
          outside any groupings, continue and all will be well.}%
 \errmessage{Input this file only once!}%
  
\fi
%
%
\def\hyperv@rsion{8}%
%
%
\newread\hyperf@le
\def\hyperf@lename{\jobname.hrf}%
\immediate\openin\hyperf@le\hyperf@lename\relax
\ifeof\hyperf@le\relax
 \immediate\closein\hyperf@le\relax
\else
 \immediate\closein\hyperf@le\relax
 \input \hyperf@lename
\fi
%
%
\newwrite\hyperf@le
\immediate\openout\hyperf@le\hyperf@lename
%
%
\newtoks\hypert@ks
%
%
\edef\hypert@mp{\catcode`\noexpand\#=\the\catcode`\#}%
\catcode`\#=12
\def\hyperh@sh{#}%
\hypert@mp
\let\hypert@mp=\relax
\let\hyper@nd=\relax
\def\hyperstr@pquote"#1"#2\hyper@nd{\ifx\hyper@ndefined#2\hyper@ndefined#1\else
                                    \ifx\hyper@ndefined#1\hyper@ndefined
                                    \hyperstr@pquote#2"\hyper@nd\else
                                    #1\hyperstr@pquote"#2"\hyper@nd\fi\fi}%
\def\hyperstr@pblank" #1 #2\hyper@nd"{\ifx\hyper@ndefined#2\hyper@ndefined#1\else
                                    \ifx\hyper@ndefined#1\hyper@ndefined
                                    \hyperstr@pblank"#2 \hyper@nd"\else
                                    #1\hyperstr@pblank" #2 \hyper@nd"\fi\fi}
\long\def\hyper@nchor#1#2{\edef\hyperm@cro{html:<A #1>}%
                          \special\expandafter{\hyperm@cro}%
                          {#2}}%
\def\hyper@atm@ning#1->#2\hyper@nd{#2}
\def\hyperlink#1{\edef\hypert@mp{#1}%
               \edef\hypert@mp{\expandafter\hyper@atm@ning\meaning\hypert@mp
                               \hyper@nd}%
               \edef\hypert@mp"{ \expandafter\hyperstr@pquote\expandafter"%
                               \hypert@mp"\hyper@nd}%
               \edef\hypert@mp{\expandafter\hyperstr@pblank\expandafter%
                               "\hypert@mp" \hyper@nd"}%
               \hyper@nchor{href=\expandafter"\hypert@mp"}}%
\def\hypertarget#1{\edef\hypert@mp{#1}%
               \edef\hypert@mp{\expandafter\hyper@atm@ning\meaning\hypert@mp
                               \hyper@nd}%
               \edef\hypert@mp"{ \expandafter\hyperstr@pquote\expandafter"%
                               \hypert@mp"\hyper@nd}%
               \edef\hypert@mp{\expandafter\hyperstr@pblank\expandafter%
                               "\hypert@mp" \hyper@nd"}%
               \hyper@nchor{name=\expandafter"\hypert@mp"}}%
\def\hyperref{\afterassignment\hyperr@f\let\hyperp@ram}
\def\hyperr@f{\ifx\hyperp@ram{\iffalse}\fi
               \expandafter\expandafter\expandafter\hyperr@@
               \expandafter{%
              \else
               \iffalse}\fi
               \ifx\hyperp@ram\hyper@ndefined
                 \message{Undefined reference}%
                 \def\hyperp@r@m{{}{undefined}{}}%
               \else
                 \edef\hyperp@r@m{\hyperp@ram}%
               \fi
               \expandafter\expandafter\expandafter\hyperr@@
               \expandafter\hyperp@r@m
              \fi}%
\def\hyperr@@#1#2#3{\ifx\hyper@ndefined#1\hyper@ndefined
                    \hypert@ks\expandafter{\hyperh@sh#2.#3}%
                    \else
                     \ifx\hyper@ndefined#2#3\hyper@ndefined
                      \hypert@ks{#1}%
                     \else
                      \def\hypert@mp{#1}%
                      \hypert@ks\expandafter\expandafter\expandafter
                      {\expandafter\hypert@mp\hyperh@sh#2.#3}%
                     \fi
                    \fi
                    \expandafter\hyperlink\expandafter{\the\hypert@ks}}%
\def\hyperdef#1#2#3{{\global\escapechar=`\\\relax
                     \edef\hypert@mp{\hyperstr@pquote"#2.#3"\hyper@nd}%
                     \expandafter\ifx\csname hyperd@\meaning\hypert@mp
                     \endcsname
                     \relax
                     \expandafter\gdef\csname hyperd@\meaning\hypert@mp
                     \endcsname{}%
                     \gdef#1{{}{\hyperstr@pquote"#2"\hyper@nd}%
                               {\hyperstr@pquote"#3"\hyper@nd}}%
                     \immediate\write\hyperf@le{\def\noexpand#1{#1}}%
                     \xdef\hypert@mp{\global\let\noexpand\hypert@mp=\relax
                                     \noexpand\hypertarget{\hypert@mp}}%
                     \global\hypert@ks={\hypert@mp}%
                     \else
                     \message\expandafter{'\hypert@mp' duplicate}%
                     \global\let\hypert@mp=\relax
                     \global\hypert@ks={\hyperdef{#1}{#2}{#3@}}%
                     \fi}\the\hypert@ks}%

\def\hyper@nique#1#2#3#4{\global\escapechar=`\\\relax
                     \edef\hypert@mp{\hyperstr@pquote"#2.#3"\hyper@nd}%
                     \expandafter\ifx\csname hyperd@\meaning\hypert@mp
                     \endcsname
                     \relax
                     \gdef#1{{}{\hyperstr@pquote"#2"\hyper@nd}%
                               {\hyperstr@pquote"#3"\hyper@nd}}%
                     \global\let\hypert@mp=\relax
                     #4%
                     \else
                     \global\let\hypert@mp=\relax
                     \hyper@nique{#1}{#2}{#3@}{#4}%
                     \fi
                     }%

\let\hyper@@@@=\relax
\def\hyper@@{\let\hyper@@@=\relax}%
\hyper@@
\def\hyper@{\relax\let\hyper@@@\noexpand\hyper@\noexpand}%
\def\hyperpr@ref{\hyper@@\hyperref}
\def\hyperpr@def{\hyper@@\hyperdef}

\let\href\hyperlink

%
%
\hypers@fe
 
%
%
\def\unredoffs{} \def\redoffs{\voffset=-.31truein\hoffset=-.48truein}
\def\speclscape{}
%
%
%
%
%
\newbox\leftpage \newdimen\fullhsize \newdimen\hstitle \newdimen\hsbody
\tolerance=1000\hfuzz=2pt
\catcode`\@=11 
\ifx\hyperdef\UNd@FiNeD\def\hyperdef#1#2#3#4{#4}\def\hyperref#1#2#3#4{#4}\fi
\def\bigans{b }
\def\answ{b }
%
\ifx\answ\bigans\message{(This will come out unreduced.}
\magnification=1200\unredoffs\baselineskip=16pt plus 2pt minus 1pt
\hsbody=\hsize \hstitle=\hsize 
\else\message{(This will be reduced.} \let\l@r=L
\magnification=1000\baselineskip=16pt plus 2pt minus 1pt \vsize=7truein
\redoffs \hstitle=8truein\hsbody=4.75truein\fullhsize=10truein\hsize=\hsbody
\output={\ifnum\pageno=0 
  \shipout\vbox{\speclscape{\hsize\fullhsize\makeheadline}
    \hbox to \fullhsize{\hfill\pagebody\hfill}}\advancepageno
  \else
  \almostshipout{\leftline{\vbox{\pagebody\makefootline}}}\advancepageno
  \fi}
\def\almostshipout#1{\if L\l@r \count1=1 \message{[\the\count0.\the\count1]}
      \global\setbox\leftpage=#1 \global\let\l@r=R
 \else \count1=2
  \shipout\vbox{\speclscape{\hsize\fullhsize\makeheadline}
      \hbox to\fullhsize{\box\leftpage\hfil#1}}  \global\let\l@r=L\fi}
\fi
%
\newcount\yearltd\yearltd=\year\advance\yearltd by -1900

\def\Title#1#2{\nopagenumbers\abstractfont\hsize=\hstitle\rightline{#1}%
\vskip 1in\centerline{\titlefont #2}\abstractfont\vskip .5in\pageno=0}
\def\Date#1{\vfill\leftline{#1}\tenpoint\supereject\global\hsize=\hsbody%
\footline={\hss\tenrm\hyperdef\hypernoname{page}\folio\folio\hss}}%
%

\def\draftmode{\message{ DRAFTMODE }\def\draftdate{{\rm preliminary draft:
\number\month/\number\day/\number\yearltd\ \ \hourmin}}%
\headline={\hfil\draftdate}\writelabels\baselineskip=20pt plus 2pt minus 2pt
 {\count255=\time\divide\count255 by 60 \xdef\hourmin{\number\count255}
  \multiply\count255 by-60\advance\count255 by\time
  \xdef\hourmin{\hourmin:\ifnum\count255<10 0\fi\the\count255}}}
\def\nolabels{\def\wrlabeL##1{}\def\eqlabeL##1{}\def\reflabeL##1{}}
\def\writelabels{\def\wrlabeL##1{\leavevmode\vadjust{\rlap{\smash%
{\line{{\escapechar=` \hfill\rlap{\sevenrm\hskip.03in\string##1}}}}}}}%
\def\eqlabeL##1{{\escapechar-1\rlap{\sevenrm\hskip.05in\string##1}}}%
\def\reflabeL##1{\noexpand\llap{\noexpand\sevenrm\string\string\string##1}}}
\nolabels
%
\global\newcount\secno \global\secno=0
\global\newcount\meqno \global\meqno=1
\def\s@csym{}
\def\newsec#1{\global\advance\secno by1%
{\toks0{#1}\message{(\the\secno. \the\toks0)}}%
\global\subsecno=0\eqnres@t\let\s@csym\secsym\xdef\secn@m{\the\secno}\noindent
{\bf\hyperdef\hypernoname{section}{\the\secno}{\the\secno.} #1}%
\writetoca{{\string\hyperref{}{section}{\the\secno}{\the\secno.}} {#1}}%
\par\nobreak\medskip\nobreak}
\def\eqnres@t{\xdef\secsym{\the\secno.}\global\meqno=1\bigbreak\bigskip}
\def\sequentialequations{\def\eqnres@t{\bigbreak}}\xdef\secsym{}
\global\newcount\subsecno \global\subsecno=0
\def\subsec#1{\global\advance\subsecno by1%
{\toks0{#1}\message{(\s@csym\the\subsecno. \the\toks0)}}%
\ifnum\lastpenalty>9000\else\bigbreak\fi
\noindent{\it\hyperdef\hypernoname{subsection}{\secn@m.\the\subsecno}%
{\secn@m.\the\subsecno.} #1}\writetoca{\string\quad
{\string\hyperref{}{subsection}{\secn@m.\the\subsecno}{\secn@m.\the\subsecno.}}
{#1}}\par\nobreak\medskip\nobreak}
\def\appendix#1#2{\global\meqno=1\global\subsecno=0\xdef\secsym{\hbox{#1.}}%
\bigbreak\bigskip\noindent{\bf Appendix \hyperdef\hypernoname{appendix}{#1}%
{#1.} #2}{\toks0{(#1. #2)}\message{\the\toks0}}%
\xdef\s@csym{#1.}\xdef\secn@m{#1}%
\writetoca{\string\hyperref{}{appendix}{#1}{Appendix {#1.}} {#2}}%
\par\nobreak\medskip\nobreak}
%
%
\def\checkm@de#1#2{\ifmmode{\def\f@rst##1{##1}\hyperdef\hypernoname{equation}%
{#1}{#2}}\else\hyperref{}{equation}{#1}{#2}\fi}
\def\eqnn#1{\DefWarn#1\xdef #1{(\noexpand\relax\noexpand\checkm@de%
{\s@csym\the\meqno}{\secsym\the\meqno})}%
\wrlabeL#1\writedef{#1\leftbracket#1}\global\advance\meqno by1}
\def\f@rst#1{\c@t#1a\em@ark}\def\c@t#1#2\em@ark{#1}
\def\eqna#1{\DefWarn#1\wrlabeL{#1$\{\}$}%
\xdef #1##1{(\noexpand\relax\noexpand\checkm@de%
{\s@csym\the\meqno\noexpand\f@rst{##1}}{\hbox{$\secsym\the\meqno##1$}})}
\writedef{#1\numbersign1\leftbracket#1{\numbersign1}}\global\advance\meqno by1}
\def\eqn#1#2{\DefWarn#1%
\xdef #1{(\noexpand\hyperref{}{equation}{\s@csym\the\meqno}%
{\secsym\the\meqno})}$$#2\eqno(\hyperdef\hypernoname{equation}%
{\s@csym\the\meqno}{\secsym\the\meqno})\eqlabeL#1$$%
\writedef{#1\leftbracket#1}\global\advance\meqno by1}
\def\xeqn{\expandafter\xe@n}\def\xe@n(#1){#1}
\def\xeqna#1{\expandafter\xe@n#1}
\def\eqns#1{(\e@ns #1{\hbox{}})}
\def\e@ns#1{\ifx\UNd@FiNeD#1\message{eqnlabel \string#1 is undefined.}%
\xdef#1{(?.?)}\fi{\let\hyperref=\relax\xdef\next{#1}}%
\ifx\next\em@rk\def\next{}\else%
\ifx\next#1\xeqn#1\else\def\n@xt{#1}\ifx\n@xt\next#1\else\xeqna#1\fi
\fi\let\next=\e@ns\fi\next}

\def\DefWarn#1{\ifx\UNd@FiNeD#1\else
\immediate\write16{*** WARNING: the label \string#1 is already defined ***}\fi}
%
\newskip\footskip\footskip14pt plus 1pt minus 1pt 
\def\footnotefont{\ninepoint}\def\f@t#1{\footnotefont #1\@foot}
\def\f@@t{\baselineskip\footskip\bgroup\footnotefont\aftergroup\@foot\let\next}
\setbox\strutbox=\hbox{\vrule height9.5pt depth4.5pt width0pt}
\global\newcount\ftno \global\ftno=0
\def\foot{\global\advance\ftno by1\def\foot@rg{\hyperref{}{footnote}%
{\the\ftno}{\the\ftno}\xdef\foot@rg{\noexpand\hyperdef\noexpand\hypernoname%
{footnote}{\the\ftno}{\the\ftno}}}\footnote{$^{\foot@rg}$}}
%
\newwrite\ftfile
\def\footend{\def\foot{\global\advance\ftno by1\chardef\wfile=\ftfile
\hyperref{}{footnote}{\the\ftno}{$^{\the\ftno}$}%
\ifnum\ftno=1\immediate\openout\ftfile=\jobname.fts\fi%
\immediate\write\ftfile{\noexpand\smallskip%
\noexpand\item{\noexpand\hyperdef\noexpand\hypernoname{footnote}
{\the\ftno}{f\the\ftno}:\ }\pctsign}\findarg}%
\def\footatend{\vfill\eject\immediate\closeout\ftfile{\parindent=20pt
\centerline{\bf Footnotes}\nobreak\bigskip\input \jobname.fts }}}
\def\footatend{}
%
%
\global\newcount\refno \global\refno=1
\newwrite\rfile
\def\ref{[\hyperref{}{reference}{\the\refno}{\the\refno}]\nref}
\def\nref#1{\DefWarn#1%
\xdef#1{[\noexpand\hyperref{}{reference}{\the\refno}{\the\refno}]}%
\writedef{#1\leftbracket#1}%
\ifnum\refno=1\immediate\openout\rfile=\jobname.refs\fi
\chardef\wfile=\rfile\immediate\write\rfile{\noexpand\item{[\noexpand\hyperdef%
\noexpand\hypernoname{reference}{\the\refno}{\the\refno}]\ }%
\reflabeL{#1\hskip.31in}\pctsign}\global\advance\refno by1\findarg}
\def\findarg#1#{\begingroup\obeylines\newlinechar=`\^^M\pass@rg}
{\obeylines\gdef\pass@rg#1{\writ@line\relax #1^^M\hbox{}^^M}%
\gdef\writ@line#1^^M{\expandafter\toks0\expandafter{\striprel@x #1}%
\edef\next{\the\toks0}\ifx\next\em@rk\let\next=\endgroup\else\ifx\next\empty%
\else\immediate\write\wfile{\the\toks0}\fi\let\next=\writ@line\fi\next\relax}}
\def\striprel@x#1{} \def\em@rk{\hbox{}}
\def\lref{\begingroup\obeylines\lr@f}
\def\lr@f#1#2{\DefWarn#1\gdef#1{\let#1=\UNd@FiNeD\ref#1{#2}}\endgroup\unskip}

\def\addref#1{\immediate\write\rfile{\noexpand\item{}#1}} 
\def\listrefs{\footatend\vfill\supereject\immediate\closeout\rfile\writestoppt
\baselineskip=\footskip\centerline{{\bf References}}\bigskip{\parindent=20pt%
\frenchspacing\escapechar=` \input \jobname.refs\vfill\eject}\nonfrenchspacing}
\def\startrefs#1{\immediate\openout\rfile=\jobname.refs\refno=#1}
\def\xref{\expandafter\xr@f}\def\xr@f[#1]{#1}
\def\refs#1{\count255=1[\r@fs #1{\hbox{}}]}
\def\r@fs#1{\ifx\UNd@FiNeD#1\message{reflabel \string#1 is undefined.}%
\nref#1{need to supply reference \string#1.}\fi%
\vphantom{\hphantom{#1}}{\let\hyperref=\relax\xdef\next{#1}}%
\ifx\next\em@rk\def\next{}%
\else\ifx\next#1\ifodd\count255\relax\xref#1\count255=0\fi%
\else#1\count255=1\fi\let\next=\r@fs\fi\next}
%

%
\newwrite\ffile\global\newcount\figno \global\figno=1
\def\fig{fig.~\hyperref{}{figure}{\the\figno}{\the\figno}\nfig}
\def\nfig#1{\DefWarn#1%
\xdef#1{fig.~\noexpand\hyperref{}{figure}{\the\figno}{\the\figno}}%
\writedef{#1\leftbracket fig.\noexpand~\xfig#1}%
\ifnum\figno=1\immediate\openout\ffile=\jobname.figs\fi\chardef\wfile=\ffile%
{\let\hyperref=\relax
\immediate\write\ffile{\noexpand\medskip\noexpand\item{Fig.\ %
\noexpand\hyperdef\noexpand\hypernoname{figure}{\the\figno}{\the\figno}. }
\reflabeL{#1\hskip.55in}\pctsign}}\global\advance\figno by1\findarg}
\def\listfigs{\vfill\eject\immediate\closeout\ffile{\parindent40pt
\baselineskip14pt\centerline{{\bf Figure Captions}}\nobreak\medskip
\escapechar=` \input \jobname.figs\vfill\eject}}
\def\xfig{\expandafter\xf@g}\def\xf@g fig.\penalty\@M\ {}
\def\figs#1{figs.~\f@gs #1{\hbox{}}}
\def\f@gs#1{{\let\hyperref=\relax\xdef\next{#1}}\ifx\next\em@rk\def\next{}\else
\ifx\next#1\xfig #1\else#1\fi\let\next=\f@gs\fi\next}
\def\figin{\epsfcheck\figin}\def\figins{\epsfcheck\figins}
\def\epsfcheck{\ifx\epsfbox\UNd@FiNeD
\message{(NO epsf.tex, FIGURES WILL BE IGNORED)}
\gdef\figin##1{\vskip2in}\gdef\figins##1{\hskip.5in}
\else\message{(FIGURES WILL BE INCLUDED)}%
\gdef\figin##1{##1}\gdef\figins##1{##1}\fi}
\def\DefWarn#1{}
\def\figinsert{\goodbreak\midinsert}
\def\ifig#1#2#3{\DefWarn#1\xdef#1{fig.~\noexpand\hyperref{}{figure}%
{\the\figno}{\the\figno}}\writedef{#1\leftbracket fig.\noexpand~\xfig#1}%
\figinsert\figin{\centerline{#3}}\medskip\centerline{\vbox{\baselineskip12pt
\advance\hsize by -1truein\noindent\wrlabeL{#1=#1}\footnotefont%
{\bf Fig.~\hyperdef\hypernoname{figure}{\the\figno}{\the\figno}:} #2}}
\bigskip\endinsert\global\advance\figno by1}
\newwrite\lfile
{\escapechar-1\xdef\pctsign{\string\%}\xdef\leftbracket{\string\{}
\xdef\rightbracket{\string\}}\xdef\numbersign{\string\#}}
\def\writedefs{\immediate\openout\lfile=\jobname.defs \def\writedef##1{%
{\let\hyperref=\relax\let\hyperdef=\relax\let\hypernoname=\relax
 \immediate\write\lfile{\string\def\string##1\rightbracket}}}}%
\def\writestop{\def\writestoppt{\immediate\write\lfile{\string\pageno
 \the\pageno\string\startrefs\leftbracket\the\refno\rightbracket
 \string\def\string\secsym\leftbracket\secsym\rightbracket
 \string\secno\the\secno\string\meqno\the\meqno}\immediate\closeout\lfile}}
\def\writestoppt{}\def\writedef#1{}
\def\seclab#1{\DefWarn#1%
\xdef #1{\noexpand\hyperref{}{section}{\the\secno}{\the\secno}}%
\writedef{#1\leftbracket#1}\wrlabeL{#1=#1}}
\def\subseclab#1{\DefWarn#1%
\xdef #1{\noexpand\hyperref{}{subsection}{\secn@m.\the\subsecno}%
{\secn@m.\the\subsecno}}\writedef{#1\leftbracket#1}\wrlabeL{#1=#1}}
\def\applab#1{\DefWarn#1%
\xdef #1{\noexpand\hyperref{}{appendix}{\secn@m}{\secn@m}}%
\writedef{#1\leftbracket#1}\wrlabeL{#1=#1}}
\newwrite\tfile \def\writetoca#1{}
\def\leaderfill{\leaders\hbox to 1em{\hss.\hss}\hfill}
\def\writetoc{\immediate\openout\tfile=\jobname.toc
   \def\writetoca##1{{\edef\next{\write\tfile{\noindent ##1
   \string\leaderfill {\string\hyperref{}{page}{\noexpand\number\pageno}%
                       {\noexpand\number\pageno}} \par}}\next}}}
\newread\ch@ckfile
\def\listtoc{\immediate\closeout\tfile\immediate\openin\ch@ckfile=\jobname.toc
\ifeof\ch@ckfile\message{no file \jobname.toc, no table of contents this pass}%
\else\closein\ch@ckfile\centerline{\bf Contents}\nobreak\medskip%
{\baselineskip=12pt\footnotefont\parskip=0pt\catcode`\@=11\input\jobname.toc
\catcode`\@=12\bigbreak\bigskip}\fi}
\catcode`\@=12 
%
\edef\tfontsize{\ifx\answ\bigans scaled\magstep3\else scaled\magstep4\fi}
\font\titlerm=cmr10 \tfontsize \font\titlerms=cmr7 \tfontsize
\font\titlermss=cmr5 \tfontsize \font\titlei=cmmi10 \tfontsize
\font\titleis=cmmi7 \tfontsize \font\titleiss=cmmi5 \tfontsize
\font\titlesy=cmsy10 \tfontsize \font\titlesys=cmsy7 \tfontsize
\font\titlesyss=cmsy5 \tfontsize \font\titleit=cmti10 \tfontsize
\skewchar\titlei='177 \skewchar\titleis='177 \skewchar\titleiss='177
\skewchar\titlesy='60 \skewchar\titlesys='60 \skewchar\titlesyss='60
\def\titlefont{\def\rm{\fam0\titlerm}
\textfont0=\titlerm \scriptfont0=\titlerms \scriptscriptfont0=\titlermss
\textfont1=\titlei \scriptfont1=\titleis \scriptscriptfont1=\titleiss
\textfont2=\titlesy \scriptfont2=\titlesys \scriptscriptfont2=\titlesyss
\textfont\itfam=\titleit \def\it{\fam\itfam\titleit}\rm}
 \ifx\answ\bigans\else scaled\magstep1\fi
\ifx\answ\bigans\def\abstractfont{\tenpoint}\else
\font\absit=cmti10 scaled \magstep1
\font\abssl=cmsl10 scaled \magstep1
\font\absrm=cmr10 scaled\magstep1 \font\absrms=cmr7 scaled\magstep1
\font\absrmss=cmr5 scaled\magstep1 \font\absi=cmmi10 scaled\magstep1
\font\absis=cmmi7 scaled\magstep1 \font\absiss=cmmi5 scaled\magstep1
\font\abssy=cmsy10 scaled\magstep1 \font\abssys=cmsy7 scaled\magstep1
\font\abssyss=cmsy5 scaled\magstep1 \font\absbf=cmbx10 scaled\magstep1
\skewchar\absi='177 \skewchar\absis='177 \skewchar\absiss='177
\skewchar\abssy='60 \skewchar\abssys='60 \skewchar\abssyss='60
\def\abstractfont{\def\rm{\fam0\absrm}
\textfont0=\absrm \scriptfont0=\absrms \scriptscriptfont0=\absrmss
\textfont1=\absi \scriptfont1=\absis \scriptscriptfont1=\absiss
\textfont2=\abssy \scriptfont2=\abssys \scriptscriptfont2=\abssyss
\textfont\itfam=\absit \def\it{\fam\itfam\absit}\def\footnotefont{\tenpoint}%
\textfont\slfam=\abssl \def\sl{\fam\slfam\abssl}%
\textfont\bffam=\absbf \def\bf{\fam\bffam\absbf}\rm}\fi
\def\tenpoint{\def\rm{\fam0\tenrm}
\textfont0=\tenrm \scriptfont0=\sevenrm \scriptscriptfont0=\fiverm
\textfont1=\teni  \scriptfont1=\seveni  \scriptscriptfont1=\fivei
\textfont2=\tensy \scriptfont2=\sevensy \scriptscriptfont2=\fivesy
\textfont\itfam=\tenit \def\it{\fam\itfam\tenit}\def\footnotefont{\ninepoint}%
\textfont\bffam=\tenbf \def\bf{\fam\bffam\tenbf}\def\sl{\fam\slfam\tensl}\rm}
\font\ninerm=cmr9 \font\sixrm=cmr6 \font\ninei=cmmi9 \font\sixi=cmmi6
\font\ninesy=cmsy9 \font\sixsy=cmsy6 \font\ninebf=cmbx9
\font\nineit=cmti9 \font\ninesl=cmsl9 \skewchar\ninei='177
\skewchar\sixi='177 \skewchar\ninesy='60 \skewchar\sixsy='60
\def\ninepoint{\def\rm{\fam0\ninerm}
\textfont0=\ninerm \scriptfont0=\sixrm \scriptscriptfont0=\fiverm
\textfont1=\ninei \scriptfont1=\sixi \scriptscriptfont1=\fivei
\textfont2=\ninesy \scriptfont2=\sixsy \scriptscriptfont2=\fivesy
\textfont\itfam=\ninei \def\it{\fam\itfam\nineit}\def\sl{\fam\slfam\ninesl}%
\textfont\bffam=\ninebf \def\bf{\fam\bffam\ninebf}\rm}
%
%

\hyphenation{anom-aly anom-alies coun-ter-term coun-ter-terms}
\def\inv{^{\raise.15ex\hbox{${\scriptscriptstyle -}$}\kern-.05em 1}}

\def\Dsl{\,\raise.15ex\hbox{/}\mkern-13.5mu D} 
\def\dsl{\raise.15ex\hbox{/}\kern-.57em\partial}

\def\lspace{\ifx\answ\bigans{}\else\qquad\fi}
\def\lbspace{\ifx\answ\bigans{}\else\hskip-.2in\fi} 
\def\boxeqn#1{\vcenter{\vbox{\hrule\hbox{\vrule\kern3pt\vbox{\kern3pt
	\hbox{${\displaystyle #1}$}\kern3pt}\kern3pt\vrule}\hrule}}}
\def\mbox#1#2{\vcenter{\hrule \hbox{\vrule height#2in
		\kern#1in \vrule} \hrule}}  
%

\def\darr#1{\raise1.5ex\hbox{$\leftrightarrow$}\mkern-16.5mu #1}

\def\half{{\textstyle{1\over2}}} 
\def\roughly#1{\raise.3ex\hbox{$#1$\kern-.75em\lower1ex\hbox{$\sim$}}}

\input epsf.tex

\lref\AM{
  L.~F.~Alday and J.~M.~Maldacena,
  JHEP {\bf 0706}, 064 (2007)
  [arXiv:0705.0303 [hep-th]].
}
\lref\MaldacenaRE{
  J.~M.~Maldacena,
  Adv.\ Theor.\ Math.\ Phys.\  {\bf 2}, 231 (1998)
  [Int.\ J.\ Theor.\ Phys.\  {\bf 38}, 1113 (1999)]
  [arXiv:hep-th/9711200].
}
\lref\WittenQJ{
  E.~Witten,
  Adv.\ Theor.\ Math.\ Phys.\  {\bf 2}, 253 (1998)
  [arXiv:hep-th/9802150].
}
\lref\GubserBC{
  S.~S.~Gubser, I.~R.~Klebanov and A.~M.~Polyakov,
  Phys.\ Lett.\  B {\bf 428}, 105 (1998)
  [arXiv:hep-th/9802109].
}
\lref\KarchSH{
  A.~Karch and E.~Katz,
  JHEP {\bf 0206}, 043 (2002)
  [arXiv:hep-th/0205236].
}
\lref\McGreevyKT{
  J.~McGreevy and A.~Sever,
  JHEP {\bf 0802}, 015 (2008)
  [arXiv:0710.0393 [hep-th]].
}
\lref\BM{
  N.~Berkovits and J.~Maldacena,
  JHEP {\bf 0809}, 062 (2008)
  [arXiv:0807.3196 [hep-th]].
}
\lref\ST{
  C.~M.~Sommerfield and C.~B.~Thorn,
  Phys.\ Rev.\  D {\bf 78}, 046005 (2008)
  [arXiv:0805.0388 [hep-th]].
}
\lref\AldayZZA{
  L.~F.~Alday and J.~Maldacena,
  AIP Conf.\ Proc.\  {\bf 1031}, 43 (2008).
}
\lref\AldayGA{
  L.~F.~Alday and J.~Maldacena,
  arXiv:0903.4707 [hep-th].
}
\lref\KomargodskiER{
  Z.~Komargodski and S.~S.~Razamat,
  JHEP {\bf 0801}, 044 (2008)
  [arXiv:0707.4367 [hep-th]].
}
\lref\McGreevyZY{
  J.~McGreevy and A.~Sever,
  JHEP {\bf 0808}, 078 (2008)
  [arXiv:0806.0668 [hep-th]].
}
\lref\BernIZ{
  Z.~Bern, L.~J.~Dixon and V.~A.~Smirnov,
  Phys.\ Rev.\  D {\bf 72}, 085001 (2005)
  [arXiv:hep-th/0505205].
}
\lref\BernKR{
  Z.~Bern, L.~J.~Dixon and D.~A.~Kosower,
  Nucl.\ Phys.\  B {\bf 412}, 751 (1994)
  [arXiv:hep-ph/9306240].
}


\lref\BrandhuberYX{
  A.~Brandhuber, P.~Heslop and G.~Travaglini,
  Nucl.\ Phys.\  B {\bf 794}, 231 (2008)
  [arXiv:0707.1153 [hep-th]].
}
\lref\McGreevyZY{
  J.~McGreevy and A.~Sever,
  JHEP {\bf 0808}, 078 (2008)
  [arXiv:0806.0668 [hep-th]].
}
\lref\AldayYW{
  L.~F.~Alday and R.~Roiban,
  Phys.\ Rept.\  {\bf 468}, 153 (2008)
  [arXiv:0807.1889 [hep-th]].
}
\lref\DixonWI{
  L.~J.~Dixon,
  arXiv:hep-ph/9601359.
}
\lref\DrummondAUA{
  J.~M.~Drummond, G.~P.~Korchemsky and E.~Sokatchev,
  Nucl.\ Phys.\  B {\bf 795}, 385 (2008)
  [arXiv:0707.0243 [hep-th]].
}
\lref\NguyenYA{
  D.~Nguyen, M.~Spradlin and A.~Volovich,
  Phys.\ Rev.\  D {\bf 77}, 025018 (2008)
    [arXiv:0709.4665$\qquad\qquad\qquad$ [hep-th]].
}
\lref\AldayYN{
  L.~F.~Alday and J.~Maldacena,
  JHEP {\bf 0911}, 082 (2009)
  [arXiv:0904.0663 [hep-th]].
}
\lref\DobashiSJ{
  S.~Dobashi and K.~Ito,
  Nucl.\ Phys.\  B {\bf 819}, 18 (2009)
  [arXiv:0901.3046 [hep-th]].
}
\lref\AstefaneseiBK{
  D.~Astefanesei, S.~Dobashi, K.~Ito and H.~Nastase,
  JHEP {\bf 0712}, 077 (2007)
  [arXiv:0710.1684 [hep-th]].
}
\lref\KorchemskyHM{
  G.~P.~Korchemsky and E.~Sokatchev,
  Nucl.\ Phys.\  B {\bf 832}, 1 (2010)
  [arXiv:0906.1737 [hep-th]].
}
\lref\DrummondFD{
  J.~M.~Drummond, J.~M.~Henn and J.~Plefka,
  JHEP {\bf 0905}, 046 (2009)
  [arXiv:0902.2987 [hep-th]].
}
\lref\BeisertIQ{
  N.~Beisert, R.~Ricci, A.~A.~Tseytlin and M.~Wolf,
  Phys.\ Rev.\  D {\bf 78}, 126004 (2008)
  [arXiv:0807.3228 [hep-th]].
}
\lref\DrummondAQ{
  J.~M.~Drummond, J.~Henn, G.~P.~Korchemsky and E.~Sokatchev,
  Nucl.\ Phys.\  B {\bf 815}, 142 (2009)
  [arXiv:0803.1466 [hep-th]].
}
\lref\DrummondAU{
  J.~M.~Drummond, J.~Henn, G.~P.~Korchemsky and E.~Sokatchev,
  Nucl.\ Phys.\  B {\bf 826}, 337 (2010)
  [arXiv:0712.1223 [hep-th]].
}
\lref\DixonGR{
  L.~J.~Dixon, L.~Magnea and G.~Sterman,
  JHEP {\bf 0808}, 022 (2008)
  [arXiv:0805.3515 [hep-ph]].
}
\lref\GardiZV{
  E.~Gardi and L.~Magnea,
  Nuovo Cim.\  {\bf 032C}, 137 (2009)
  [arXiv:0908.3273 [hep-ph]].
}

\lref\DixonUR{
  L.~J.~Dixon, E.~Gardi and L.~Magnea,
  JHEP {\bf 1002}, 081 (2010)
  [arXiv:0910.3653 [hep-ph]].
}

\lref\ItoZY{
  K.~Ito, H.~Nastase and K.~Iwasaki,
  Prog.\ Theor.\ Phys.\  {\bf 120}, 99 (2008)
  [arXiv:0711.3532 [hep-th]].
}
\lref\AldayZM{
  L.~F.~Alday, J.~M.~Henn, J.~Plefka and T.~Schuster,
  JHEP {\bf 1001}, 077 (2010)
  [arXiv:0908.0684 [hep-th]].
}
\lref\HennBK{
  J.~M.~Henn, S.~G.~Naculich, H.~J.~Schnitzer and M.~Spradlin,
  JHEP {\bf 1004}, 038 (2010)
  [arXiv:1001.1358 [hep-th]].
}
\lref\GrossKZA{
       D.~J.~Gross and P.~F.~Mende,
       Phys.\ Lett.\ B {\bf 197}, 129 (1987).
     }
\lref\GrossGE{
 D.~J.~Gross and J.~L.~Manes,
 Nucl.\ Phys.\ B {\bf 326}, 73 (1989).
}
\lref\CollinsBT{
  J.~C.~Collins,
  Adv.\ Ser.\ Direct.\ High Energy Phys.\  {\bf 5}, 573 (1989)
  [arXiv:hep-ph/0312336].
}
\lref\CataniEF{
  S.~Catani, S.~Dittmaier and Z.~Trocsanyi,
  Phys.\ Lett.\  B {\bf 500}, 149 (2001)
  [arXiv:hep-ph/0011222].
}
\lref\KruczenskiME{
  M.~Kruczenski, L.~A.~P.~Zayas, J.~Sonnenschein and D.~Vaman,
  JHEP {\bf 0506}, 046 (2005)
  [arXiv:hep-th/0410035].
}

\Title{\vbox{\baselineskip12pt }}
{\vbox{\centerline{Massive Quark Scattering at Strong Coupling from AdS/CFT}}}
\centerline{ Edwin Barnes and Diana Vaman}
\bigskip
\centerline{Department of Physics} \centerline{University of
Virginia} \centerline{Charlottesville, VA 22904, USA}
\centerline{Email addresses: efbarnes, dv3h@virginia.edu}
\bigskip
\noindent

We extend the analysis of Alday and Maldacena for obtaining gluon scattering amplitudes at strong coupling to include external massive quark states. Our quarks are actually the ${\cal N}{=}2$ hypermultiplets which arise when $D7$-brane probes are included in the $AdS_5\times S^5$ geometry. We work in the quenched approximation, treating the ${\cal N}{=}2$ matter multiplets as external sources coupled to the ${\cal N}{=}4$ SYM fields. We first derive appropriate massive-particle boundary conditions for the string scattering worldsheets. We then find an exact worldsheet which corresponds to the scattering of two massive quarks and two massless gluons and extract from this the associated scattering amplitude. We also find the worldsheet and amplitude for the scattering of four massive quarks. Our worldsheet solutions reduce to the four massless gluon solution of Alday and Maldacena in the limit of zero quark mass. The amplitudes we compute can also be interpreted in terms of 2-2 scattering involving gluons and massive W-bosons.

\Date{October 2009}

\newsec{Introduction}

As shown by Alday and Maldacena \AM, planar scattering amplitudes in ${\cal N}{=}4$ supersymmetric Yang-Mills theory at strong coupling can be obtained from AdS/CFT \MaldacenaRE\WittenQJ\GubserBC\ by computing the area of a string worldsheet in AdS whose boundary conditions are dictated by the kinematic data of the in-going and out-going scattering states. More precisely, the string theory calculation computes a kinematic factor in the color-ordered\foot{See \DixonWI\ for a review on the color decomposition of QCD amplitudes.} amplitude of the form

\eqn\amplitudedecayfactor{{\cal A}\sim e^{i S_{cl}}=e^{-{\sqrt{\lambda}\over2\pi}\times\hbox{area}}.}
$S_{cl}$ represents the classical string action, and ``area" refers to the area of the classical string worldsheet exhibiting the appropriate boundary conditions. The 't Hooft coupling $\lambda$ appears as an overall factor multiplying the string action. The fact that it is large in AdS/CFT applications ($\sqrt{\lambda}\sim(\alpha')^{-1}\to\infty$) allows one to perform the saddlepoint approximation leading to expression \amplitudedecayfactor. In their original paper, Alday and Maldacena (AM) focused on four-gluon scattering, and at least for this case, they argue that the string action $S_{cl}$ should be taken to be precisely the Nambu-Goto action (i.e. without any additional boundary terms) when expressed in terms of Poincar\'{e} coordinates on the ``T-dual" $AdS_5$. In other words, the worldsheet area in \amplitudedecayfactor\ is not the area as measured in the original $AdS$, but rather in a different $AdS$ obtained by performing a particular coordinate transformation on the Minkowski subspace of the original $AdS$. This ``T-duality" is so named because of its mathematical resemblance to the usual T-duality, but it is performed on non-compact dimensions.

The original work of AM has been extended along a
number of directions, including gluon amplitudes
with more than four external legs \AldayGA\AldayYN\DobashiSJ
\AstefaneseiBK, amplitudes with massless external quarks
\McGreevyKT\KomargodskiER, and finite temperature scattering
\ItoZY. It has also led to new insights into the structure of
${\cal N}{=}4$ SYM amplitudes. The authors of \DrummondAUA\ showed
that these amplitudes obey a hidden symmetry which they refer to
as `dual conformal symmetry', and further evidence for this symmetry has since been accrued \KorchemskyHM\BM\DrummondFD\BeisertIQ\NguyenYA.
Along with this discovery came indications of a connection between
${\cal N}{=}4$ scattering amplitudes and Wilson loops
\DrummondAUA\BM\DrummondAU\McGreevyZY\DrummondAQ\BrandhuberYX. The work of AM has also led to new insights
into the structure of IR divergences in gluon scattering amplitudes \DixonGR\GardiZV\DixonUR. Much of this progress is reviewed in \AldayYW.

In this paper, we will generalize the work of AM to the case in which either two or all four of the external lines are massive quarks as opposed to gluons. We will modify the gravity setup of AM by including extra probe D7-branes as in the work of Karch and Katz \KarchSH, so that external ${\cal N}{=}2$ hypermultiplets will play the role of dynamical quarks, and it is more appropriate to think of the amplitudes we obtain as being field theory amplitudes in the quenched approximation. This approach was also taken by McGreevy and Sever in \McGreevyKT\ and by Komargodski and Razamat in \KomargodskiER. In both cases, only massless quarks were considered.

In the case of massless quarks, McGreevy and Sever provided some evidence that there exist certain singularities (folded strings) on the T-dual worldsheet boundary. This evidence arises by conjecturing that the massless quark-gluon worldsheet can be obtained as half of a six-gluon worldsheet with self-crossing, where the point of self-crossing is proposed to imitate the effect of Neumann boundary conditions on a D7-brane, which in turn allow for the possibility of folded-string singularities in the worldsheet boundary. These folded strings, they argue, prevent the external massless quark states from giving rise to the Sudakov-like factors \CollinsBT\ that are present for external gluons. However, a mechanism which determines the extent of folded strings into the $AdS$ bulk remains unknown and, in particular, it is not clear if these folded strings would be stable and have finite extent. McGreevy and Sever further give an analysis of planar diagrams to argue that field theory expectations corroborate the absence of Sudakov factors for external massless quarks, thus motivating the folded-string conjecture.

In this paper, we do not include folded strings for two reasons. First, in the case of massive quarks (which is the primary focus of this paper), the arguments of McGreevy and Sever do not necessarily apply. Indeed as we will show, it is not necessary to invoke singularities on the worldsheet boundary in order to eliminate quark Sudakov factors---the non-zero mass alone removes these factors. Secondly,
in other problems involving string configurations ending on branes (e.g. mesons of large spin which are described by spinning open strings ending on D-branes), the leading alpha-prime contribution comes from strings which end at the boundary of the D-brane. The alpha-prime perturbative expansion will include contributions which come from string fluctuations in those directions which have Neumann boundary conditions on the D-brane, or in other words, from fluctuations of the string endpoint inside the D-brane worldvolume. However, these contributions are sub-leading \KruczenskiME. 


Our results may also be interpreted as scattering amplitudes involving gluons and massive W-bosons. In this case, folded strings are definitely absent since Dirichlet boundary conditions are imposed on both ends of the W-boson strings. Recently, it has been shown that perturbative scattering amplitudes in Higgsed ${\cal N}{=}4$ SYM exhibit dual conformal symmetry when this symmetry is appropriately extended to incorporate the mass of the gauge bosons \AldayZM\HennBK. This result strongly suggests that the amplitude/Wilson loop correspondence extends to massive W-bosons as well. In light of the relationship between Wilson loops and scattering amplitudes in ${\cal N}{=}4$ SYM, it is not surprising that W-bosons and ${\cal N}{=}2$ hypermultiplets share the same kinematic factor \amplitudedecayfactor\ since these representations are treated the same way in Wilson loop computations. In this paper, we will always refer to the massive states as quarks, leaving implicit this additional interpretation.


The methods we employ to obtain the kinematic factor \amplitudedecayfactor\ in massive quark-gluon and quark-quark scattering amplitudes are interesting in their own right. We show quite generally how worldsheets for massive-state scattering can be obtained from those pertaining to massless states. It is likely that these techniques may be applied to study other types of scattering processes.

The first step in constructing worldsheets which represent massive quark scattering is to identify the right boundary conditions for these worldsheets. This is done in a straight-forward manner by first finding the extended open string solutions which should correspond to massive quarks moving at subluminal velocities. Applying the same ``T-duality" as in AM then leads to the desired boundary conditions in the dual $AdS_5$. The result is that the worldsheet boundary condition corresponding to a massive quark turns out to be a line segment extending from a point on the boundary of the T-dual $AdS$ to a point in the bulk. This line segment is light-like in terms of the full five-dimensional $AdS$.

The next step is to find the classical worldsheet solutions which obey these boundary conditions. This process is greatly facilitated by making use of a solution-generating trick first described in \AM. This trick enables one to find worldsheets representing four-particle amplitudes starting from worldsheets ending on a single light-like cusp. The latter have a boundary behavior which is sufficiently simple that the worldsheets can be found easily by exploiting Poincar\'{e} invariance. The trick then entails performing a particular set of rotations in global $AdS$ coordinates on the single-cusp solutions. AM made use of this trick to find the four-gluon ($gggg$) worldsheet by starting from a single light-like cusp worldsheet for which the cusp lay at the boundary of the T-dual $AdS$. We will show that by starting with single-cusp solutions with the cusp lying a finite distance away from the boundary, one can use the same trick to obtain worldsheets corresponding to the scattering of two massive and two massless particles ($ggqq$) (see Figs. 3 and 4) and of four massive particles ($qqqq$) (see Figs. 8 and 9). This method yields $ggqq$ amplitudes in which both quarks have the same mass and $qqqq$ amplitudes in which two quarks have the same mass, and the other two share a different mass. It is not obvious how to extend our approach to more general particle masses, and we leave this as an open problem.

Technically, the S-matrix for quark/gluon scattering at strong coupling is ill defined due to confinement, but in practice, it is still useful to compute such amplitudes as intermediate steps in the computation of real physical observables. Quark/gluon amplitudes reflect the fact that they are unphysical by exhibiting IR divergences. As shown by AM, this behavior is borne out by the string theory calculation, which yields the amplitude predicted by Bern, Dixon and Smirnov \BernIZ\ only after the string worldsheet area has been properly regularized. In their original paper, AM employed dimensional regularization to this end. In \AldayZZA, they also showed that one can obtain the same field theory amplitude kinematic factor by making use of a radial cutoff regularization scheme for the worldsheet area.

In this paper, we will also make use of the radial cutoff scheme in the computation of the kinematic factors in the $ggqq$ and $qqqq$ amplitudes. We find that the leading IR divergence of the $ggqq$ amplitude with massive quarks is exactly half that of the four-gluon amplitude (see eqn. 4.30). It is interesting to note that the Sudakov form factor computed in perturbative QCD for massive/massless parton scattering exhibits similar behavior \CataniEF. We are also able to obtain momentum and mass-dependent terms in the amplitude, and we find that it reduces to the four-gluon kinematic factor in the small mass limit. (Note that the full amplitude need not have a smooth limit, and indeed it is only the leading terms which behave smoothly.)

In the case of the massive $qqqq$ amplitude, we find that there is an IR divergence despite the fact that all external masses are non-zero (see eqn. 5.23), and we regulate this using the radial cutoff scheme. The $qqqq$ amplitude exhibits a weaker divergence compared with that of the $ggqq$ and $gggg$ amplitudes: one obtains $\log(\hbox{IR cutoff})$ instead of $\log^2(\hbox{IR cutoff})$. This single-log behavior also arises in perturbative QCD calculations \CataniEF.


The paper is organized as follows. In section 2, we derive the worldsheet boundary conditions for massive scattering states. Section 3 is devoted to showing how one can obtain the boundary of a four-particle worldsheet from a light-like cusp lying in the $AdS$ bulk at a fixed value of the $AdS$ radial coordinate. In particular, we show that the cusp maps (under the AM trick) into the boundary of a $ggqq$ worldsheet. The solution for a worldsheet ending on a single cusp in the bulk was found in \BM. We then employ the same map to turn this solution into an exact $ggqq$ worldsheet in section 4. We plug this worldsheet into the Nambu-Goto action to obtain the kinematic factor in the amplitude. In section 5, we generalize these ideas and show that if we start with two parallel light-like cusps in the $AdS$ bulk, the same tricks turn this configuration into the $qqqq$ boundary. Next, we find the worldsheet stretching between the two light-like cusps and map this to produce the exact $qqqq$ worldsheet, from which we then extract the desired amplitude factor. Appendices A and B contain a detailed derivation and analysis of the single- and two-light-like cusp worldsheet solutions. In appendix C, we compute four-point scattering amplitudes for extended open strings in flat space to make a comparison with the results for string scattering in $AdS$.

\newsec{Worldsheet boundary conditions for massive particles}

We begin with the following form of the $AdS_5\times S^5$ metric:

\eqn\nn{ds^2=r^2dx_\mu dx^\mu+{1\over r^2}dr^2+d\Omega_5^2,}
where the $AdS_5$ factor has the mostly $+$ signature and we set the $AdS$ radius to 1. We will choose $x^8$ and $x^9$ to be the directions in which the D7-brane is localized. By combining the radial direction of $AdS_5$ with the $S^5$, we may reparametrize the metric according to

\eqn\nn{ds^2=[\rho^2+(x^8)^2+(x^9)^2]dx_\mu dx^\mu+{1\over \rho^2+(x^8)^2+(x^9)^2}[d\rho^2+\rho^2 d\Omega_3^2+(dx^8)^2+(dx^9)^2],}
with

\eqn\nn{r^2=\rho^2+(x^8)^2+(x^9)^2.}

We are interested in string configurations in which the string stretches from the stack of $D3$-branes to the probe $D7$-brane and moves with constant velocity in the four Minkowski directions. In terms of the $AdS_5$ radial coordinate $r$, the strings stretch from $r=0$ up to $r=r_{D7}$, where $r_{D7}$ denotes the radius at which the $D7$-brane appears to ``vanish in thin air" \KarchSH. Such solutions should correspond to asymptotic scattering states under AdS/CFT, with the mass of the scattered particles related to the finite extent of the strings. Without loss of generality, we may restrict attention to solutions which have the following form:

\eqn\noname{x^\mu=k^\mu\tau,\qquad x^9=0,\qquad x^8=x,\qquad \rho=\rho(x),}
with the remaining angular variables set to zero. $\tau$ can be thought of as a time direction on the string worldsheet, and the four-momentum $k^\mu$ is such that $k^2=-m^2$ for some mass $m$. $k^\mu$ and $m$ can be interpreted as the momentum and mass of an external quark state. Solutions of the above form correspond to strings which are stretched in the $x^8$ and $\rho$ directions. That is, we assume the $D7$-brane probe is separated from the stack of $D3$-branes only along the $x^8$ direction, and we allow for the possibility that the string extends in the $\rho$ direction since this direction is Neumann at the $D7$-brane. $x_{D7}$ will denote the location of the $D7$-brane in the $x^8$ direction. Plugging this ansatz into the Nambu-Goto action gives

\eqn\noname{S_{NG} = {m\over2\pi\alpha'}\int dx d\tau \sqrt{1+\rho'^2}.}
The solution to the equation of motion for $\rho(x)$ is just

\eqn\noname{\rho=\rho_1x+\rho_0.}
Imposing a Neumann boundary condition on $\rho$ at $x=x_{D7}$ requires that $\rho_1=0$, so that $\rho=\rho_0$ for all $x$. Therefore, our classical string configurations do not bend but simply stretch between the $D3$ and $D7$-branes despite the Neumann boundary condition on $\rho$. For simplicity, we will set $\rho=0$ and $x^9=0$ for the remainder of the paper and only consider worldsheets that obey these constraints. Therefore, $x_{D7}=r_{D7}$.

In accordance with the AdS/CFT correspondence, we expect the leading-order strong-coupling scattering amplitude of massive particles to be related to the area of a string worldsheet in $AdS_5$ which tends to the string configurations found above---namely straight strings stretching between the $D3$ and $D7$-branes and moving with constant velocity---at large distances in the 4d spacetime which acts as the boundary of $AdS_5$. In order to facilitate the imposition of these boundary conditions, we follow the lead of AM and ``T-dualize" these asymptotic string solutions:

\eqn\AMTdualmap{\partial_\alpha y^\mu=r^2\epsilon_{\alpha\beta}\partial_\beta x^\mu.}
We do not have the $i$ on the right-hand side as in AM because we choose our worldsheets to have Minkowski signature. The map \AMTdualmap\ is not generally covariant, however, and it is not valid if we want to work with the space-like worldsheet coordinate $x$. A natural generalization which is covariant is

\eqn\gencovTduality{\partial_\alpha y^\mu=r^2 J_\alpha^\beta \partial_\beta x^\mu,}
where $J_\alpha^\beta$ is the induced complex structure on the worldsheet. The form of the map \AMTdualmap\ then holds in the special case where the coordinate system is such that the induced metric is conformally flat. For worldsheet coordinates $\tau$ and $x$, we have

\eqn\noname{J_x^\tau={g_{xx}\over\sqrt{-g}}={1\over mx^2}.}
Plugging this into \gencovTduality, we find

\eqn\bcs{\partial_xy^\mu={k^\mu\over m}\quad\Rightarrow\quad y^\mu={k^\mu\over m}x+y_0^\mu.}
$y_0^\mu$ are integration constants. As one might expect from the case of four gluon scattering studied by AM, the asymptotic string states T-dualize to line segments which are oriented along the directions of the four-momenta of the scattered particles. The new feature here is that these line segments are now time-like (from the point of view of the $y^\mu$ subspace) and extend into the bulk of (the T-dual) $AdS_5$. This last statement is perhaps made clearer by inverting the boundary condition \bcs:

\eqn\bcforx{x=-{k_\mu(y^\mu-y_0^\mu)\over m},}
and recalling that the radial direction of $AdS_5$ is given by $r=|x|$. We see that, on the boundary of the worldsheet, $r$ varies linearly with the T-dual coordinates $y^\mu$.
\bcs\ also implies that

\eqn\deltaydeltax{\Delta y^\mu={\Delta x\over m}k^\mu,}
where $\Delta y^\mu$ and $\Delta x$ represent the amount by which $y^\mu$ and $x$ change over a segment of the boundary. Defining the quark mass to be the energy of a static string stretching from $x=0$ to $x=x_{D7}$, we have

\eqn\noname{m={|\Delta x|\over2\pi\alpha'}={x_{D7}\over2\pi\alpha'},}
so that

\eqn\noname{\Delta y^\mu=2\pi\alpha'k^\mu.}

Our boundary condition \bcs\ can be restated in a more illuminating form which is obtained by squaring both sides of \deltaydeltax:

\eqn\lightlikecondition{(\Delta y^\mu)^2+(\Delta x)^2=0.}
This is simply the statement that the bounding line segment is light-like in terms of the full five-dimensional space. From the point of view of the four-dimensional Minkowski space, we have a time-like segment with $|\Delta x|$ acting as a mass.

We observe in passing that if we wish to reproduce the massless particle boundary conditions from \bcs, we need to replace $\tau$ and $x$ with more suitable worldsheet coordinates since these become singular in the massless limit. If we first define new worldsheet coordinates $\tilde\tau\equiv \tau/m$ and $\tilde \sigma=x/m$, we can then safely take the zero mass limit in \bcs\ to reproduce the result that the worldsheet boundary is comprised of light-like segments oriented along the directions of their associated null four-momenta. Alternatively, we can start directly from \lightlikecondition, which works equally well in the case of massless particles, for which $\Delta x=0$.

In the next section, we will focus on finding string worldsheet solutions which end on four line segments, two of which are massive and two massless. These boundary conditions will therefore lead to a scattering amplitude involving two quarks and two gluons. We will first argue that four-cusp solutions of this type can be obtained from certain single-cusp solutions in much the same way that the four-cusp gluon scattering worldsheet can be obtained from the single-cusp worldsheet as shown by Alday and Maldacena \AM. We will then apply this reasoning to a particular single-cusp solution to obtain a worldsheet for $ggqq$ scattering.

In section 5, we will look for solutions ending on four massive line segments. Such solutions will lead to the kinematic factor for four-quark scattering. These solutions will be obtained in a manner similar to the $ggqq$ solutions, but we will start from two-cusp solutions instead of single-cusp solutions.

\newsec{Four-cusp solutions from single-cusps lying in the $AdS$ bulk}

We begin by recalling how one can obtain the four-cusp $gggg$ scattering worldsheet from a single-cusp solution \AM. The trick is to exploit the global symmetries of the full $AdS$. The worldsheet ending on a single-cusp formed by two light-like lines intersecting at the origin and lying in the $x=0$\foot{Recall that under the T-duality of AM, the $AdS$ horizon and boundary effectively swap locations, so that the boundary is now at $r=0$. Since $\rho=x^9=0$ and since we will only be interested in non-negative values of $x$, we simply set $r=x$.} plane is given by

\eqn\AMsinglecusp{x=\sqrt{2(y_0^2-y_1^2)}.}
Making use of the usual map between Poincar\'{e} coordinates and embedding coordinates,

\eqn\poincareembeddingrel{\eqalign{Y^\mu &={y^\mu\over r},\qquad \mu=0,...,3\cr Y_{-1}+Y_4 &= {1\over r},\cr Y_{-1}-Y_4 &= {r^2+y_\mu y^\mu\over r},}}
where the $Y$'s satisfy the embedding equation

\eqn\Yembed{-Y_{-1}^2-Y_0^2+Y_1^2+Y_2^2+Y_3^2+Y_4^2=-1,}
the single-cusp solution \AMsinglecusp\ can be expressed as

\eqn\AMsinglecuspembed{Y_0^2-Y_1^2=Y_{-1}^2-Y_4^2,\qquad Y_2=Y_3=0.}
It turns out to be useful to transform this solution with the following set of rotations:

\eqn\AMtransform{Y_0\to{Y_0+Y_{-1}\over\sqrt{2}},\quad Y_{-1}\to{Y_0-Y_{-1}\over\sqrt{2}},\quad Y_1\to{Y_1+Y_2\over\sqrt{2}},\quad Y_4\to{Y_1-Y_2\over\sqrt{2}},\quad Y_2\to Y_4,}
after which the single-cusp solution \AMsinglecuspembed\ becomes

\eqn\AMsinglecuspembedii{Y_1Y_2=Y_0Y_{-1},\qquad Y_3=Y_4=0.}
The utility of the rotations \AMtransform\ is revealed when we now switch back to Poincar\'{e} coordinates to find

\eqn\AMfourcusp{y_0=y_1y_2,\qquad x=\sqrt{(1-y_1^2)(1-y_2^2)}.}
These equations describe a worldsheet which ends on four light-like line segments which join together to form a square in the $y_1/y_2$ plane described by the lines $y_1=\pm1$, $y_2=\pm1$. After regularization, the area of this worldsheet gives the leading-order (in the strong coupling limit) planar contribution to the four-gluon scattering amplitude in a special kinematic regime where the $s$ and $t$ Mandelstam variables are equal. A more general kinematic configuration can be acquired by performing an additional boost in the embedding coordinates as described in \AM.

What we have just seen is that when we lift the worldsheet solution \AMsinglecusp\ to the embedding coordinate form \AMsinglecuspembed, we pick up additional worldsheet boundary lines which lie outside the original Poincar\'{e} patch. The original two boundary lines of the single cusp intersect these additional lines at the edge of the Poincar\'{e} patch. In performing the set of rotations \AMtransform, we bring the additional boundaries into the same Poincar\'{e} patch as the original boundaries to arrive at a four-cusp worldsheet boundary.

In light of this, a natural question is what happens when our single cusp lies not at $x=0$, but instead on the $x=\epsilon$ plane. In order to understand the effect of the rotations \AMtransform\ in this case, we will first focus on the behavior of the cusp boundary lines and postpone an analysis of how the full worldsheet transforms to section 4. In Poincar\'{e} coordinates, we can describe these lines according to

\eqn\nn{x=\epsilon,\qquad y_0^2=y_1^2,\qquad y_2=y_3=0.}
These lines are plotted in figure 1.

\bigskip
\centerline{\epsfxsize=0.40\hsize\epsfbox{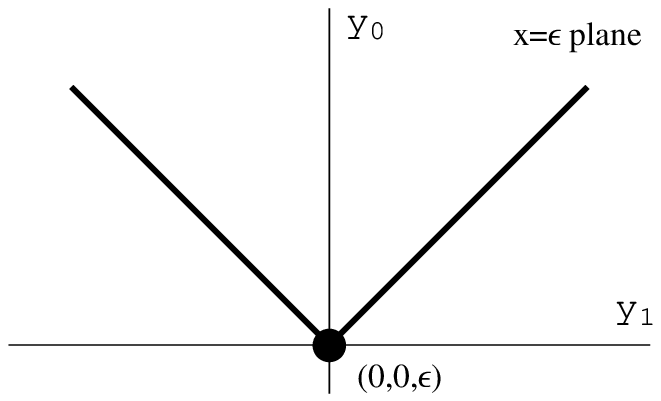}}
\centerline{\ninepoint\sl \baselineskip=2pt {\bf Figure 1:}
{\sl The bounding light-like line segments of the single-cusp }}
\centerline{\ninepoint\sl worldsheet located at $x=\epsilon$ viewed in the $y_0,y_1,x$ subspace.}
\bigskip

In terms of embedding coordinates, we can rewrite the boundary lines as

\eqn\IRbranemasslessbdy{Y_4+Y_{-1}=1/\epsilon,\qquad Y_0^2-Y_1^2=0,\qquad Y_2=Y_3=0.}
Under transformation \AMtransform, these equations become

\eqn\AMtransformbdy{Y_1-Y_2+Y_0-Y_{-1}={\sqrt{2}\over\epsilon},\qquad Y_1Y_2-Y_0Y_{-1}={1\over2},\qquad Y_3=Y_4=0,}
and re-expressing these results in terms of Poincar\'{e} coordinates yields

\eqn\poincarebdy{y_1-y_2+y_0-1={\sqrt{2}\over\epsilon}x,\qquad y_1y_2-y_0={x^2\over2},\qquad x=\sqrt{1+y_0^2-y_1^2-y_2^2},\qquad y_3=0.}
Solving for $x^2$ in the last two equations, equating the results and solving for $y_0$ gives

\eqn\yzero{y_0=|y_1+y_2|-1.}
If we now combine the first and third equations in \poincarebdy\ and solve for $y_1$ or $y_2$, we see that the transformed worldsheet boundary in the $y_1/y_2$ plane is described by the following set of four lines:

\eqn\ggqqbdy{y_1=1,\qquad y_2=-1,\qquad y_2=1-\epsilon^2+\epsilon^2y_1,\qquad y_1=-1+\epsilon^2+\epsilon^2y_2.}
The behavior of $y_0$ and $x$ along each of these line segments can easily be determined from the expressions given in \poincarebdy\ and \yzero. In the $\epsilon\to0$ limit, the lines given in \ggqqbdy\ of course become the AM four-cusp boundary $y_1=\pm1$, $y_2=\pm1$. The four cusps are located at

\eqn\noname{(y_1,y_2)=\left\{(1,1),(1,-1),(-1,-1),{1-\epsilon^2\over1+\epsilon^2}(-1,1)\right\}.}
$x$ vanishes at the first three of these points, but is nonzero at the final cusp:

\eqn\noname{x\left({\epsilon^2-1\over1+\epsilon^2},{1-\epsilon^2\over1+\epsilon^2}\right)=-2\sqrt{2}{\epsilon\over1+\epsilon^2}.}
Since $x<0$ at this cusp, we see that the boundary does not lie entirely within the Poincar\'{e} patch except in the limit $\epsilon\to0$.\foot{Recall that, even though $x=x^8\in(-\infty,\infty)$, we have assumed that $r=x\ge0$ when we employed the Poincar\'{e} embedding relations \poincareembeddingrel. Thus, $x<0$ implies that we have left the Poincar\'{e} patch.} So it appears that we fail to obtain a four-cusp boundary for finite $\epsilon$.

However, a minor adjustment to the preceding analysis will cure this problem and lead us to a four-cusp boundary. After performing the rotations \AMtransform, we should look at a Poincar\'{e} patch different from the one given in \poincareembeddingrel. That is, starting from \AMtransformbdy, we look at the Poincar\'{e} patch defined by the relations

\eqn\poincareembeddingrelii{\eqalign{Y^\mu &=-{y^\mu\over r},\qquad \mu=0,...,3\cr Y_{-1}+Y_4 &= -{1\over r},\cr Y_{-1}-Y_4 &= -{r^2+y_\mu y^\mu\over r}.}}
In this patch, the solution looks like

\eqn\poincarebdyii{y_1-y_2+y_0-1=-{\sqrt{2}\over\epsilon}x,\qquad y_1y_2-y_0={x^2\over2},\qquad x=\sqrt{1+y_0^2-y_1^2-y_2^2},\qquad y_3=0.}
Compared with \poincarebdy, we have effectively taken $\epsilon\to-\epsilon$ so that the fourth cusp is now located at

\eqn\nn{x\left({\epsilon^2-1\over1+\epsilon^2},{1-\epsilon^2\over1+\epsilon^2}\right)=2\sqrt{2}{\epsilon\over1+\epsilon^2}\equiv x_{D7}>0,}
and the problem we encountered before no longer arises. The final transformed boundary lines are plotted in figure 2.

\bigskip
\centerline{\epsfxsize=0.40\hsize\epsfbox{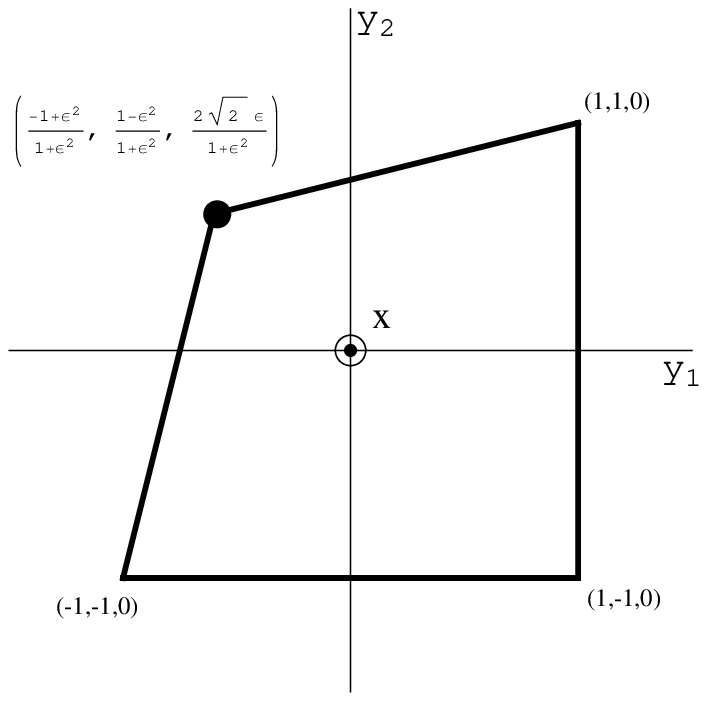}}
\centerline{\ninepoint\sl \baselineskip=2pt {\bf Figure 2:}
{\sl The bounding time-like line segments of the four-cusp $ggqq$ worldsheet viewed}}
\centerline{\ninepoint\sl in the $y_1,y_2,x$ subspace. The cusp with the black dot is the image of the cusp in figure 1.}
\bigskip

To confirm that this boundary would describe $ggqq$ scattering, we must verify that the two segments which intersect at the $x=x_{D7}$ cusp obey the boundary conditions we derived earlier (see equation \bcs). In view of \lightlikecondition, it suffices to check that the change in $x$ along these line segments, namely $x_{D7}$, is equal to the proper length of the segments as measured in the Minkowski subspace parametrized by $y_\mu$. Consider the segment which extends from the point ${1-\epsilon^2\over1+\epsilon^2}(-1,1)$ to the point $(1,1)$. Along this segment, the coordinates change by the following amounts:

\eqn\noname{\Delta y_0=2,\qquad \Delta y_1={2\over1+\epsilon^2},\qquad \Delta y_2={2\epsilon^2\over1+\epsilon^2}.}
Thus we find

\eqn\noname{-(\Delta y_\mu)^2=(\Delta y_0)^2-(\Delta y_1)^2-(\Delta y_2)^2={8\epsilon^2\over(1+\epsilon^2)^2}=x_{D7}^2=(\Delta x)^2.}
The same analysis applied to the segment extending from the point ${1-\epsilon^2\over1+\epsilon^2}(-1,1)$ to the point $(-1,-1)$ reveals that this segment has the same proper length and thus obeys \lightlikecondition\ as well. This fact is not surprising since the original segment was light-like, and the conformal transformation we performed should preserve this property from the five-dimensional point of view.

We should also point out that the $ggqq$ worldsheet boundary lines derived in this section correspond to a special kinematic regime in which the Mandelstam variables $s$ and $t$ are related to each other and are of order one, since both will be functions of $\epsilon$ only. To consider more general configurations, it is necessary to perform additional global coordinate transformations. For example, to change the overall scale of the momenta, we can perform a boost in the $Y_{-1}/Y_4$ plane. This will simply rescale the entire worldsheet while leaving the shape in-tact. If we wish to change the ratio $s/t$, then we must alter the relative orientations of the four worldsheet boundary lines. As shown in \AM, this can be accomplished by doing a boost in the $Y_0/Y_4$ plane. Parametrizing the boost by $\beta$ such that

\eqn\noname{Y_0\to\sqrt{1+\beta^2}Y_0-\beta Y_4, \qquad Y_4\to\sqrt{1+\beta^2}Y_4-\beta Y_0,}
one can show that the boundary lines \ggqqbdy\ become

\eqn\ggqqbdyii{\eqalign{y_1&=1-\beta y_2,\quad y_2=-1-\beta y_1,\quad (1-\beta\epsilon^2)y_2=1-\epsilon^2+(\epsilon^2-\beta)y_1,\cr &(1-\beta\epsilon^2)y_1=-1+\epsilon^2+(\epsilon^2-\beta)y_2.}}
The expression \yzero\ for $y_0$ along the boundary generalizes to

\eqn\yzeroii{y_0={\sqrt{1+\beta^2}\over1-\beta}(|y_1+y_2|-1),}
and $x$ is now given by

\eqn\noname{x=\sqrt{1-{2\beta\over\sqrt{1+\beta^2}}y_0+y_0^2-y_1^2-y_2^2}.}

We will now compute the Mandelstam variables associated with this modified worldsheet boundary. Recalling that each particle four-momentum is given by the change in $y^\mu$ over each boundary segment, $2\pi\alpha' k_i^\mu=\Delta y^\mu$, we obtain the following momenta:

\eqn\momenta{\eqalign{2\pi\alpha' k_1^\mu&={2\over1-\beta^2}\left(\sqrt{1+\beta^2},-1,\beta\right),\cr 2\pi\alpha' k_2^\mu &= {2\over1-\beta^2}\left(-\sqrt{1+\beta^2},{\epsilon^2-\beta\over1+\epsilon^2},{1-\beta\epsilon^2\over1+\epsilon^2}\right),\cr 2\pi\alpha' k_3^\mu &= {2\over1-\beta^2}\left(\sqrt{1+\beta^2},{1-\beta\epsilon^2\over1+\epsilon^2},{\epsilon^2-\beta\over1+\epsilon^2}\right),\cr 2\pi\alpha' k_4^\mu &= {2\over1-\beta^2}\left(-\sqrt{1+\beta^2},\beta,-1\right).}}
The corresponding masses are

\eqn\masses{m_1^2=m_4^2=0,\qquad m_2^2=m_3^2={8\epsilon^2\over(2\pi\alpha')^2(1-\beta)^2(1+\epsilon^2)^2}.}
The Mandelstam variables $s$ and $t$ are given by

\eqn\mandelstam{s=-(k_1+k_2)^2=-{8\over(2\pi\alpha')^2(1-\beta)^2(1+\epsilon^2)^2},\qquad t=-(k_1+k_4)^2=-{8\over(2\pi\alpha')^2(1+\beta)^2}.}
From \masses\ and \mandelstam, it is evident that we can obtain any ratio of $s/t$ by varying $\beta\in(-1,1)$. (It can be checked that values of $\beta$ outside this range lead to negative values of $x$.) The masses $m_2=m_3$ can be held fixed while $s/t$ is varied by choosing $\epsilon(\beta)$ appropriately. $s$, $t$, and $m_2$ can all be made independent by performing an additional scale transformation on all the coordinates as discussed above. Also note that we may keep the masses and Mandelstam variables finite in the $\alpha'\to0$ limit by rescaling the T-dual Poincar\'{e} coordinates by $\alpha'$. Of course, the final amplitudes will only depend on ratios of these quantities and would be unaffected by such a rescaling.

In summary, we have shown that one can obtain the boundary of a $ggqq$ scattering worldsheet by applying appropriate transformations to a single light-like cusp boundary located at a fixed value of the radial coordinate in the $AdS$ bulk. Thus, one would expect that if we start with a worldsheet solution which ends on the single cusp, then under the same set of transformations, this will become a worldsheet solution ending on the $ggqq$ boundary. Since these transformations are $SO(2,4)$ rotations under which the action is invariant, the worldsheet obtained from the mapping is guaranteed to be a solution. Fortunately, worldsheets ending on single cusps which lie in the $AdS$ bulk have been found already by Berkovits and Maldacena \BM. In the next section, we will make use of their solutions to show that one can indeed find a $ggqq$ worldsheet in this fashion.

\newsec{A worldsheet for $ggqq$ scattering}

In appendix B of \BM, Berkovits and Maldacena (BM) wrote down a one-parameter family of solutions for worldsheets ending on a light-like cusp lying on an ``IR regulator" brane, that is a $D3$-brane in the bulk of $AdS$. In appendix A, we review this class of solutions and argue that only one of these is relevant for the purpose of obtaining a $ggqq$ scattering worldsheet. This solution was also of particular interest to BM as it is the only solution which reduces to the single-cusp solution \AMsinglecusp\ in the limit that the regulator brane rejoins the rest of the $D3$-brane stack, i.e. when $\epsilon\to0$. That is, this solution can be thought of as a regularized version of \AMsinglecusp. It can be obtained by expressing the Lagrangian in terms of hyperbolic coordinates $w$ and $\sigma$:

\eqn\hyperboliccoords{y_0=e^{\tau(w)}\cosh\sigma,\qquad y_1=e^{\tau(w)}\sinh\sigma.}
As discussed in detail in the appendix, the $w$ and $\sigma$ coordinates are especially useful because it is possible to find solutions for which $x$ is a function of $w$ only and which obey the boundary conditions at $x=\epsilon$. Indeed, this fact is crucial to obtaining all the analytic solutions of this paper. This fact also leads to a relatively simple expression for the Lagrangian as we will see shortly. $x$ is given by the ansatz

\eqn\noname{x(w)=we^{\tau(w)},}
and the equation of motion for $\tau(w)$ is readily solved with the result

\eqn\tauiii{\tau(w)=-\log(w+1)-{1\over\sqrt{2}}\log(w-\sqrt{2})+{1\over\sqrt{2}}\log(w+\sqrt{2})+\log\epsilon.}
Returning to Poincar\'{e} coordinates, this solution becomes

\eqn\BMsinglecusp{x+\sqrt{y_0^2-y_1^2}=\epsilon\left(x+\sqrt{2(y_0^2-y_1^2)}\over x-\sqrt{2(y_0^2-y_1^2)}\right)^{1/\sqrt{2}},\qquad y_2=y_3=0.}
It is easy to see that this solution ends on the light-like cusp on the regulator brane, i.e. it contains the lines $y_0^2=y_1^2$, $x=\epsilon$. This solution gives $x$ as an implicit function of $y_0$ and $y_1$. Applying the $SO(2,4)$ rotations described in the previous section, we arrive at the four-cusp worldsheet:

\eqn\ggqqws{-{\epsilon\over\sqrt{2}}{y_1-y_2+y_0-1\over\sqrt{1+y_0^2-y_1^2-y_2^2}}=f\left(\sqrt{{1\over2}+{y_0-y_1y_2\over1+y_0^2-y_1^2-y_2^2}}\right),\quad x=\sqrt{1+y_0^2-y_1^2-y_2^2},}
where we have defined

\eqn\noname{f(z)\equiv (1+z)\left({1-\sqrt{2}z\over1+\sqrt{2}z}\right)^{1/\sqrt{2}},}
and as usual, $y_3$ remains zero after the conformal transformations. A plot of the worldsheet is shown from different viewpoints in figures 3 and 4.

\bigskip
$$\matrix{\epsfxsize=0.40\hsize\epsfbox{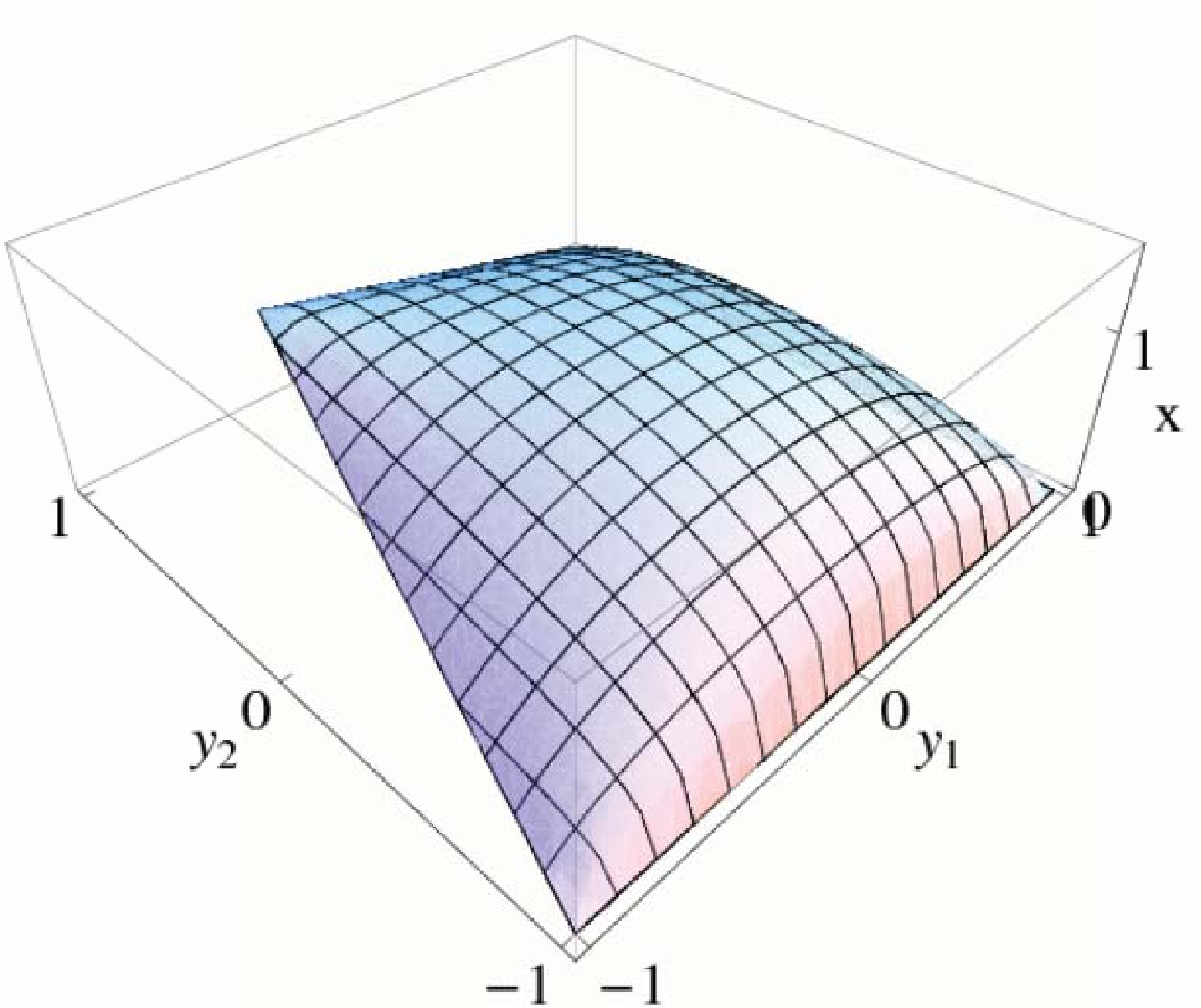} & \qquad & \epsfxsize=0.3\hsize\epsfbox{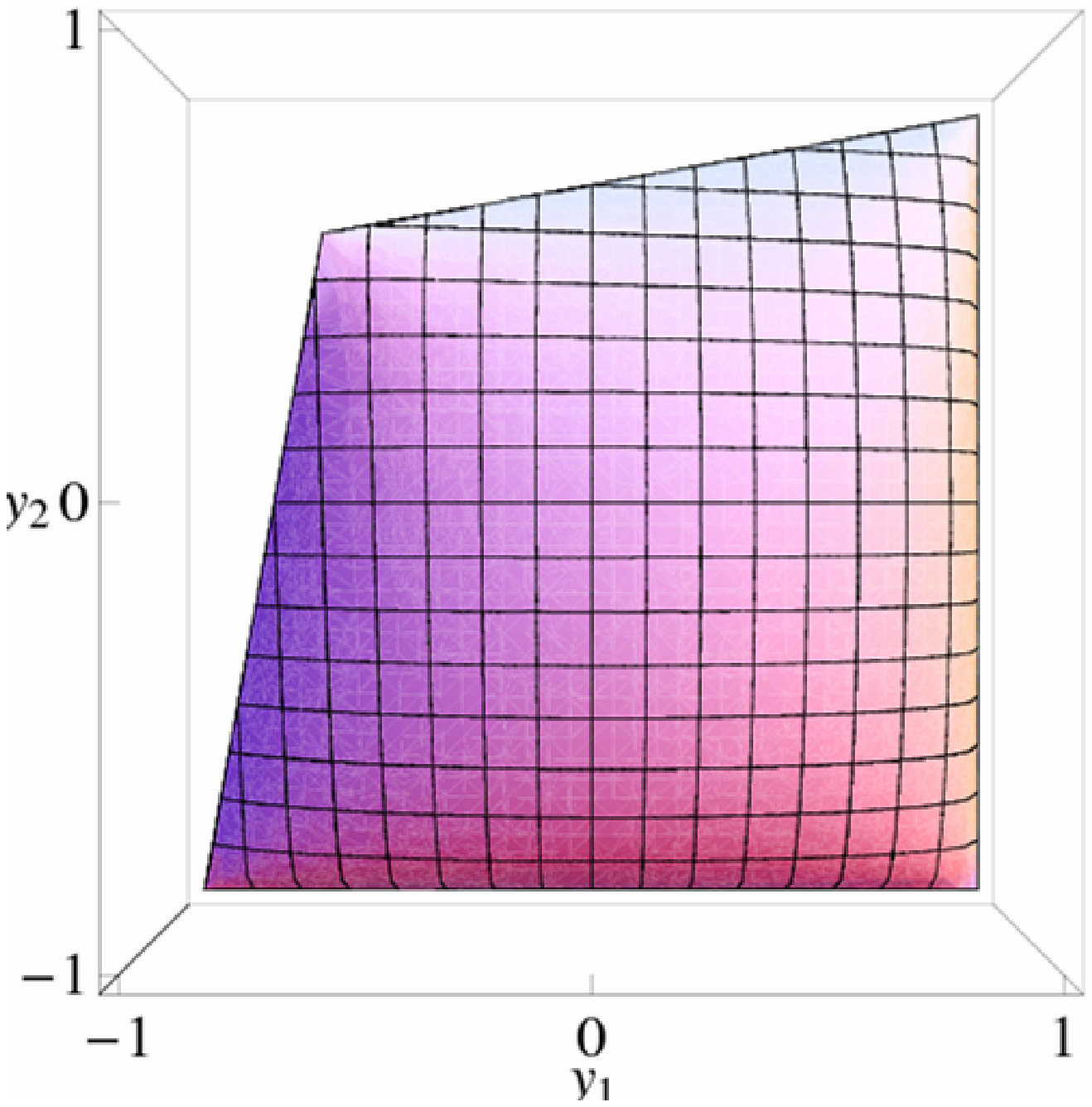}}$$
\centerline{\ninepoint\sl \baselineskip=2pt {\bf Figure 3:}
{\sl Side and top views of $x(y_1,y_2)$ for the transformed BM worldsheet with $\epsilon=1/2$.}}
\bigskip

\bigskip
$$\matrix{\epsfxsize=0.40\hsize\epsfbox{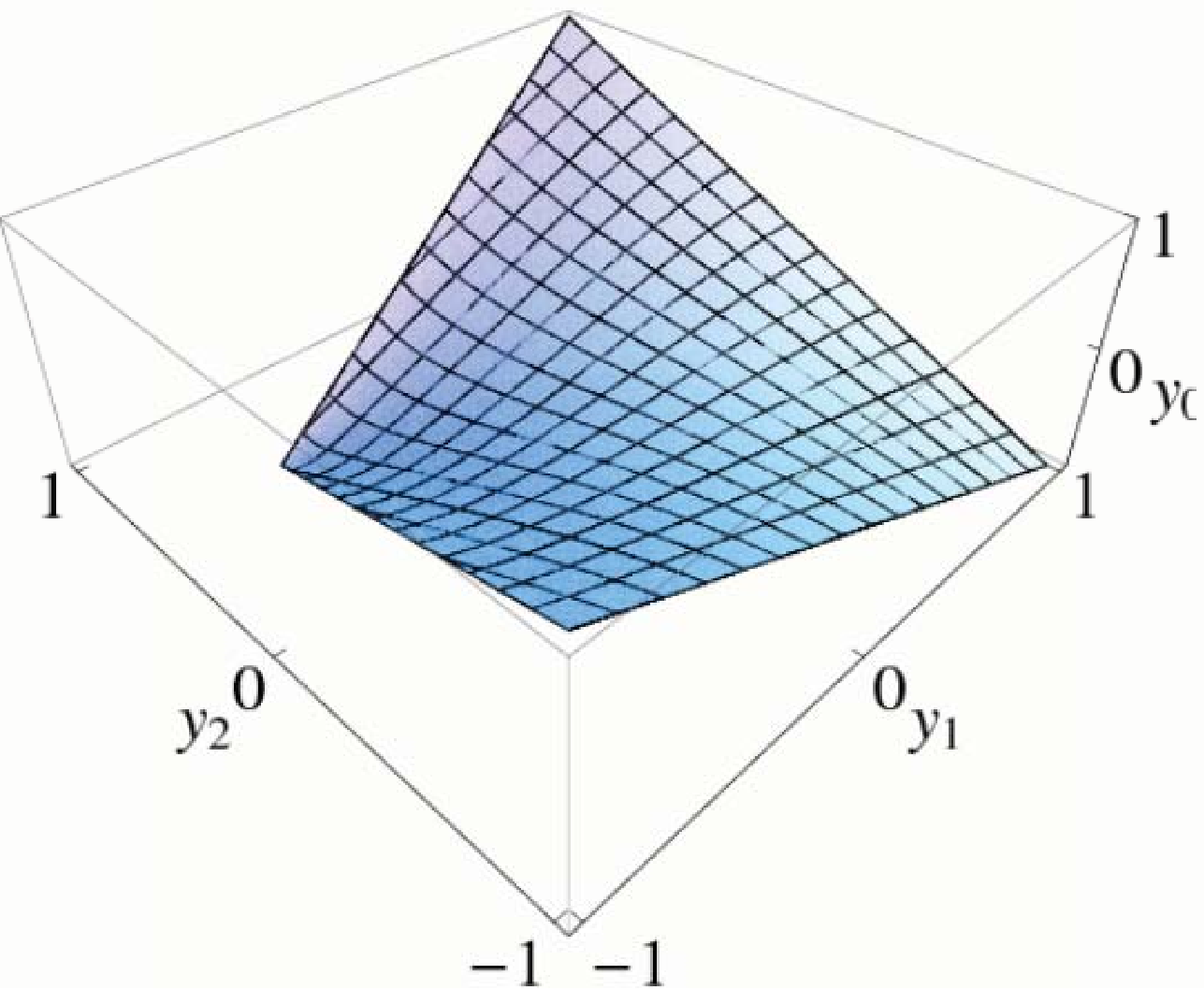} & \qquad & \epsfxsize=0.3\hsize\epsfbox{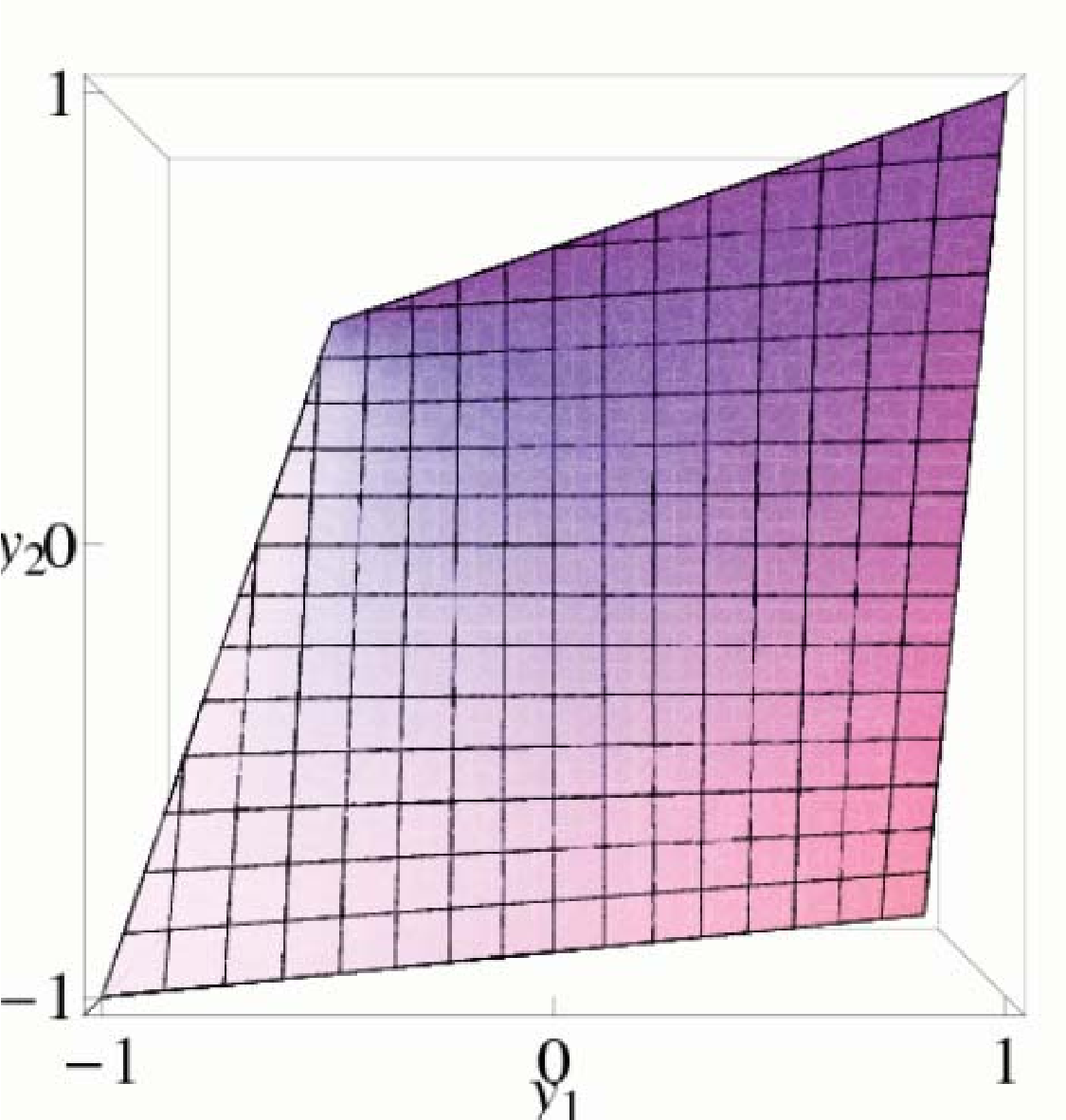}}$$
\centerline{\ninepoint\sl \baselineskip=2pt {\bf Figure 4:}
{\sl Side and top views of $y_0(y_1,y_2)$ for the transformed BM worldsheet with $\epsilon=1/2$.}}
\bigskip

\noindent Figures 3 and 4 clearly exhibit the correct qualitative behavior near the boundary. That it also agrees quantitatively can be checked directly from \ggqqws, but this is essentially true by construction. In order to reproduce the four-gluon worldsheet, we send $\epsilon\to0$. From \ggqqws, it is clear that the solution tends to $y_0=y_1y_2$ in this limit, which is precisely the AM solution.

Next, we turn to computing the area associated with this worldsheet. The most straight-forward way to do this would be to insert this solution into the Nambu-Goto Lagrangian,

\eqn\NGlag{{\cal L}={1\over2\pi\alpha'}{1\over x^2}\sqrt{1+(\partial_i x)^2-(\partial_iy_0)^2-(\partial_1x\partial_2y_0-\partial_2x\partial_1y_0)^2},}
and integrate. In \NGlag, $i=1,2$ and $\partial_i={\partial\over\partial y_i}$. However, the fact that the solution \ggqqws\ gives only an implicit function for $y_0(y_1,y_2)$ poses a challenge since it is not obvious how to write down the Lagrangian as a function of $y_1$ and $y_2$. We can avoid this problem by describing the worldsheet parametrically in terms of coordinates on the original cusp. In particular, we may use \AMtransform\ to construct the following map between the Poincar\'{e} coordinates \ST:

\eqn\poincaremapping{\eqalign{x&={2\sqrt{2}x'\over(1-\beta)(1+x'^2+y'_\mu y'^\mu)+2(1+\beta)y'_0},\cr y_0&=\sqrt{1+\beta^2}{2y_0'-1-x'^2-y'_\mu y'^\mu\over(1-\beta)(1+x'^2+y'_\mu y'^\mu)+2(1+\beta)y'_0},\cr y_1&={-1+x'^2+y'_\mu y'^\mu+2y'_1\over(1-\beta)(1+x'^2+y'_\mu y'^\mu)+2(1+\beta)y'_0},\cr y_2&={1-x'^2-y'_\mu y'^\mu+2y'_1\over (1-\beta)(1+x'^2+y'_\mu y'^\mu)+2(1+\beta)y'_0}.}}
The primed coordinates are the old coordinates, and the unprimed are the new. Switching to hyperbolic coordinates and plugging in the solution $x'=we^{\tau(w)}$, where $\tau(w)$ is as given in \tauiii, the $ggqq$ worldsheet is described parametrically according to

\eqn\wsigmatopoincare{\eqalign{x&={2\sqrt{2}we^\tau\over(1-\beta)\left[1+(w^2-1)e^{2\tau}\right]+2(1+\beta)e^\tau\cosh\sigma},\cr y_0&=\sqrt{1+\beta^2}{2e^\tau\cosh\sigma-1-(w^2-1)e^{2\tau}\over(1-\beta)\left[1+(w^2-1)e^{2\tau}\right]+2(1+\beta)e^\tau\cosh\sigma},\cr y_1&= {-1+(w^2-1)e^{2\tau}+2e^\tau\sinh\sigma\over(1-\beta)\left[1+(w^2-1)e^{2\tau}\right]+2(1+\beta)e^\tau\cosh\sigma},\cr y_2&= {1-(w^2-1)e^{2\tau}+2e^\tau\sinh\sigma\over(1-\beta)\left[1+(w^2-1)e^{2\tau}\right]+2(1+\beta)e^\tau\cosh\sigma}.}}
The $ggqq$ worldsheet is swept out as $w$ and $\sigma$ run over the ranges $(\sqrt{2},\infty)$ and $(-\infty,\infty)$ respectively. This parametric description is probably the easiest way to generate plots like those shown in figures 3 and 4.

The Nambu-Goto action in the $w$, $\sigma$ coordinates takes the form

\eqn\actionii{S={i\over\pi\alpha'}\int_{-\infty}^\infty d\sigma\int_{\sqrt{2}}^\infty {dw\over w^2(w^2-2)}.}
This is of course identical to the single-cusp action, which is studied in the appendix. The action is clearly divergent, which is to be expected since the corresponding field theory amplitude is IR divergent and in need of regularization. In the appendix, the divergence is regulated using a UV momentum cutoff for the $y_0'$, $y_1'$ coordinates. In the present context, however, we are recasting the single-cusp solution as a four-cusp solution, and it is more appropriate to introduce a regulator in the new Poincar\'{e} coordinates $y_\mu$, $x$, since it is these coordinates that have the interpretation of momenta and an energy scale. We choose to employ the radial cutoff regularization scheme, which has been successfully applied to the case of four-gluon scattering in \AldayZZA. Since the divergence arises only from the region of the worldsheet which lies at the boundary of the T-dual $AdS$, it can be regulated by cutting off the integration at a small value of $x$ which we will call $x_c$.

Before proceeding with the implementation of this regularization scheme, we would first like to set up the calculation in such a way that we may touch base with the massless limit. Since our worldsheet reduces to the AM worldsheet in the massless limit, it should be possible to obtain the $gggg$ amplitude kinematic factor in this limit. In this vein, the $w$ coordinate poses an obstacle since it is not a good coordinate for describing the AM worldsheet solution, for which $w$ is a constant. (In particular, $w=\sqrt{2}$.) Therefore, if we wish to find a result for the action which contains both the $gggg$ and $ggqq$ kinematic factors as limiting cases, we should replace $w$ with a more suitable coordinate, and in light of the AM solution, a natural replacement is $\tau$. Using \tauiii, one finds that the action \actionii\ in terms of $\tau$ is

\eqn\actioniii{i2\pi\alpha'S=2\int_{-\infty}^\infty d\sigma\int_{-\infty}^\infty d\tau {w(\tau)+1\over w(\tau)^3[w(\tau)+2]}.}

We are now ready to implement the radial cutoff regularization scheme. From \wsigmatopoincare, we see that the curve for which $x=x_c$ can be described by $\sigma=\Sigma(\tau)$, where $\Sigma(\tau)$ is given by

\eqn\sigmaoftau{\Sigma(\tau)=\cosh^{-1}\left\{{\sqrt{2}w\over(1+\beta)x_c}-{1-\beta\over2(1+\beta)}\left[e^{-\tau}-e^\tau+w^2 e^\tau\right]\right\}.}
The regulated action is

\eqn\actioniv{i2\pi\alpha'S=4\int_{-\infty}^{\tau_c} d\tau {w+1\over w^3(w+2)}\Sigma(\tau).}
The upper integration limit $\tau_c$ is defined by the condition $\Sigma(\tau_c)\equiv0$. The $\tau$ integration limits in \actioniv\ lead to the $ggqq$ action. For the $gggg$ action, the lower limit must be replaced with $-\tau_c$, as we can see by inspecting the argument of the arccosh in \sigmaoftau, which is an even function of $\tau$ when $w=\sqrt{2}$.  We will return to this later on. Finding a closed-form expression for this integral is a daunting task considering that $w(\tau)$ is only defined implicitly through \tauiii. This is unnecessary, however, as it is still possible to extract the most important pieces of the kinematic factor in the amplitude from \actioniv\ by approximating the integrand. We will make two approximations. In the first approximation, we replace the exact form for $w(\tau)$ (which in principle is obtained by inverting $\tauiii$) with an approximate form that captures its asymptotic behavior. $w$ has the properties $w\approx\epsilon e^{-\tau}$ as $\tau\to-\infty$ and $w\to\sqrt{2}$ as $\tau\to\infty$. Furthermore, as $\epsilon\to0$, we approach the AM solution, for which $w=\sqrt{2}$ for all $\tau$. (These properties can be derived from \tauiii, and a more complete discussion is given in appendix A.) Therefore, we will exchange the exact $w(\tau)$ with the piecewise function

\eqn\piecewisew{w(\tau)=\bigg\{\matrix{\sqrt{2} & \tau\ge\log{\epsilon\over\sqrt{2}} \cr \epsilon e^{-\tau} & \tau\le\log{\epsilon\over\sqrt{2}}}.}
For any $\epsilon$, there is always a finite separation between \piecewisew\ and the exact $w(\tau)$ in the neighborhood of the point $\tau=\log{\epsilon\over\sqrt{2}}$. Despite this discrepancy, \piecewisew\ is still a very useful approximation. We have checked numerically that the effect of using \piecewisew\ instead of the exact $w(\tau)$ is to shift the value of the action by a term of the form $\log(x_c)+const.$ This term is independent of $\epsilon$. Therefore in this approximation, we cannot evaluate precisely the single $\log(x_c)$ terms in the action. We also cannot reproduce the overall additive constant, but this is scheme-dependent anyway.

We can write the action as the sum of two pieces: $S=S_{<}+S_{>}$, where $S_{<}$ is obtained by integrating the integrand of \actioniv\ over the range $\tau\in(-\infty,\log{\epsilon\over\sqrt{2}}]$, and $S_{>}$ is obtained by integrating over $\tau\in[\log{\epsilon\over\sqrt{2}},\tau_c]$. Solving $\Sigma(\tau_c)=0$ for $\tau_c$ gives

\eqn\noname{\tau_c= \cosh^{-1}\left(A-1\over B\right),}
with $A$ and $B$ defined by

\eqn\noname{A\equiv {2\over(1+\beta)x_c},\qquad B\equiv {1-\beta\over1+\beta}.}

The integral for $S_{<}$ is easily done, and the result is of the form $\log(x_c)+const.$ Since our approximation does not accurately reproduce such terms, it is not sensible to keep the contribution from $S_{<}$, and we focus solely on $S_{>}$:

\eqn\noname{-i2\pi\alpha'S_{>}=\int_{\log{\epsilon\over\sqrt{2}}}^{\tau_c}d\tau \cosh^{-1}\left(A-B\cosh\tau\right)\approx \int_{\log{\epsilon\over\sqrt{2}}}^{\tau_c}d\tau \log\left(2A-Be^{|\tau|}\right).}
Here, we have introduced our second approximation to the integrand, which assumes that $A\gg1$, which is equivalent to $x_c\ll1$. In this approximation, we are merely throwing away terms that vanish in the $x_c\to0$ limit, so the final result for the regulated action is unaffected.

It is possible to compute the indefinite integral. For $\tau$ positive, we have

\eqn\indefiniteintegralp{\int d\tau \log\left(2A-Be^\tau\right)=\tau\log(2A)-\hbox{Li}_2\left({B\over2A}e^\tau\right)+\hbox{Li}_2\left(B\over2A\right),}
while for negative $\tau$, we get

\eqn\indefiniteintegraln{\int d\tau \log\left(2A-Be^{-\tau}\right)=-{\tau^2\over2}-i\pi\tau+\tau\log B-\hbox{Li}_2\left({2A\over B}e^\tau\right)+\hbox{Li}_2\left(2A\over B\right).}
In each case, we have chosen the integration constant to be such that the antiderivative vanishes at $\tau=0$. This will ensure that the antiderivative is continuous at $\tau=0$. We then obtain for $S_{>}$,

\eqn\actionii{\eqalign{i2\pi\alpha'S_{>}&\approx\log(2A)\log\left({2A\over B}\right)-{\pi^2\over6}+{1\over2}\log^2\left(\epsilon\over\sqrt{2}\right)-\log\left(\epsilon\over\sqrt{2}\right)\log B\cr &+i\pi\log\left(\epsilon\over\sqrt{2}\right)-\hbox{Li}_2\left(2A\over B\right)+\hbox{Li}_2\left({2A\over B}{\epsilon\over\sqrt{2}}\right).}}
Also in the $A\gg1$ limit, we may approximate the dilogarithm:

\eqn\dilogapprox{\hbox{Li}_2\left(2A\over B\right)\approx {\pi^2\over3}-{1\over2}\log^2\left(2A\over B\right)-i\pi\log\left(2A\over B\right),}
Note that we have not said anything about the relationship between $A$ and $\epsilon$.  Therefore, \actionii\ should give either the $gggg$ amplitude or the $ggqq$ amplitude depending on the size of $\epsilon$ relative to $x_c$. If $\epsilon\ll x_c$, then the entire worldsheet boundary lies below the infrared cutoff, which is equivalent to considering the four-gluon case. The opposite limit gives $ggqq$ scattering. In order to obtain the $gggg$ amplitude, we should make the lower integration limit $-\tau_c$ instead of $\log{\epsilon\over\sqrt{2}}$, as the contribution from this portion of the boundary is controlled by the infrared cutoff scale instead of the mass scale in this case. In other words, we should choose

\eqn\noname{\epsilon=\sqrt{2}e^{-\cosh^{-1}[(A-1)/B]}.}
We then find in the large $A$ limit

\eqn\noname{-i2\pi\alpha'S_{gggg}\approx 2\log(2A)\log\left({2A\over B}\right)-{\pi^2\over3}.}
$A$ and $B$ can be related to the quark mass and Mandelstam variables as they are defined in \masses\ and \mandelstam:

\eqn\noname{A={\sqrt{-t}\over2\mu},\qquad B={\sqrt{st}\over m^2-s}.}
Here, the quark mass is $m=m_2=m_3$, and $\mu$ is an IR cutoff scale defined by

\eqn\noname{\mu\equiv {x_c\over2\sqrt{2}\pi\alpha'}.}
In the limit $m\to0$, we then have

\eqn\Sggggi{-i2\pi\alpha'S_{gggg}\approx 2\log\left(\mu\over\sqrt{-t}\right)\log\left(\mu\over \sqrt{-s}\right)+...}
We have replaced the $-\pi^2/3$ term with ``..." to reflect the fact that we cannot claim to have computed the correct constant piece in $S_{gggg}$, and we may also be missing terms linear in $\log(\mu)$. Setting $1/\alpha'=\sqrt{\lambda}$ where $\lambda$ is the 't Hooft coupling, and re-arranging the logarithms in \Sggggi\ yields the familiar result for the four-gluon scattering kinematic factor:

\eqn\Sggggii{-iS_{gggg}={\sqrt{\lambda}\over2\pi}\left[\log^2{\mu\over\sqrt{-s}}+\log^2{\mu\over\sqrt{-t}}-{1\over4}\log^2{s\over t}\right]+...}

Now we turn to computing $S_{ggqq}$. If we instead suppose that $\epsilon\gg x_c$, then $A\gg 1/\epsilon$, and the argument of the second dilog in \actionii\ is large. We can therefore expand it according to \dilogapprox. Expanding both dilogarithms in \actionii\ and canceling terms leaves us with

\eqn\noname{-i2\pi\alpha'S_{ggqq}\approx \log(2A)\log\left(2A\over B\right)-{\pi^2\over6}-\log(2A)\log\left(\epsilon\over\sqrt{2}\right).}
In terms of the mass and Mandelstam variables, this is

\eqn\noname{-i2\pi\alpha'S_{ggqq}\approx \log\left(\mu\over\sqrt{-t}\right)\left[\log\left(\mu\sqrt{-s}\over m^2-s\right)+\log\left(m\over\sqrt{-s}\right)\right]+...}
Here, we have used $\epsilon=m/\sqrt{-s}$, which can be derived from \masses\ and \mandelstam. Note that we have thrown away a term of the form $-\half\log{2}\log(\mu/\sqrt{-t})$ as well as the constant $-\pi^2/6$, in accordance with the validity of our approximation scheme. Replacing $\alpha'$ in favor of the 't Hooft coupling and re-arranging terms gives

\eqn\ggqqamplitude{-iS_{ggqq}\approx {\sqrt{\lambda}\over4\pi}\left[\log^2{\mu\over\sqrt{-t}}+\log^2{\mu\over\sqrt{-s}}-{1\over4}\log^2{s\over t}+2\log{\mu\over\sqrt{-t}}\log{m\sqrt{-s}\over m^2-s}\right]+...}
It is interesting to note that the leading $\log^2\mu$ divergence is half that of the four-gluon result. The same feature arises in the one-loop Sudakov form factor for 2-2 massive quark-gluon scattering in real QCD \CataniEF. Also notice that the four-gluon result can be obtained from $S_{ggqq}$ by sending $m\to\mu$. In general, one does not expect this limit to be smooth however \BernKR, and the connection between $S_{ggqq}$ and $S_{gggg}$ only appears smooth here because we have neglected subleading terms.

\bigskip

\newsec{A worldsheet for four-quark scattering}

We have seen that by generalizing the single-cusp solution of Alday and Maldacena \AMsinglecusp\ to a single-cusp solution which ends at an arbitrary value of $x$ (which we called $\epsilon$), we obtain after the transformation \AMtransform\ a four-cusp solution in which two of the bounding line segments have been given a mass. We would now like to address the question of whether or not there exists a further generalization of the AM solution that will produce, upon application of \AMtransform, a four-cusp solution where all four bounding line segments receive a mass. In what follows, we will see that there indeed exists such a generalization.

In order to figure out what the appropriate generalization is, it helps to take a closer look at how the case of $ggqq$ scattering worked. In particular, a careful inspection of the above results for $ggqq$ scattering worldsheets reveals that a 4d light-like cusp at $x=\epsilon$ transforms to a 4d time-like cusp at $x=x_{D7}$ whose boundary is specified by the following two lines:

\eqn\masscuspbdy{y_2=1-\epsilon^2+\epsilon^2y_1,\qquad y_1=-1+\epsilon^2+\epsilon^2y_2.}
The point of this cusp is located at $(y_1,y_2)={1-\epsilon^2\over1+\epsilon^2}(-1,1)$. More precisely, the original light-like cusp connects to three other cusps at infinity. Under the transformation \AMtransform, the light-like cusp becomes the time-like cusp at $x=x_{D7}$, and the other three cusps (which lie at $x=0$) come in from infinity. So there is a real sense in which the portion of the $ggqq$ boundary which lies above the line $y_1=y_2$ (also the portion for which $x\ne0$) is the image of the original light-like cusp under the map \AMtransform. We will refer to this portion of the $ggqq$ worldsheet as a ``massive" cusp.

The boundary of a $qqqq$ worldsheet would have two such massive cusps. For the case where all four massive particles are of the same mass, we can construct this boundary by reflecting the two lines in \masscuspbdy\ through the $y_1=y_2$ plane. The `mirror image' cusp then has boundary

\eqn\imagebdy{y_1=1-\epsilon^2+\epsilon^2y_2,\qquad y_2=-1+\epsilon^2+\epsilon^2y_1,}
with the tip at $(y_1,y_2)={1-\epsilon^2\over1+\epsilon^2}(1,-1)$.

\bigskip
\centerline{\epsfxsize=0.40\hsize\epsfbox{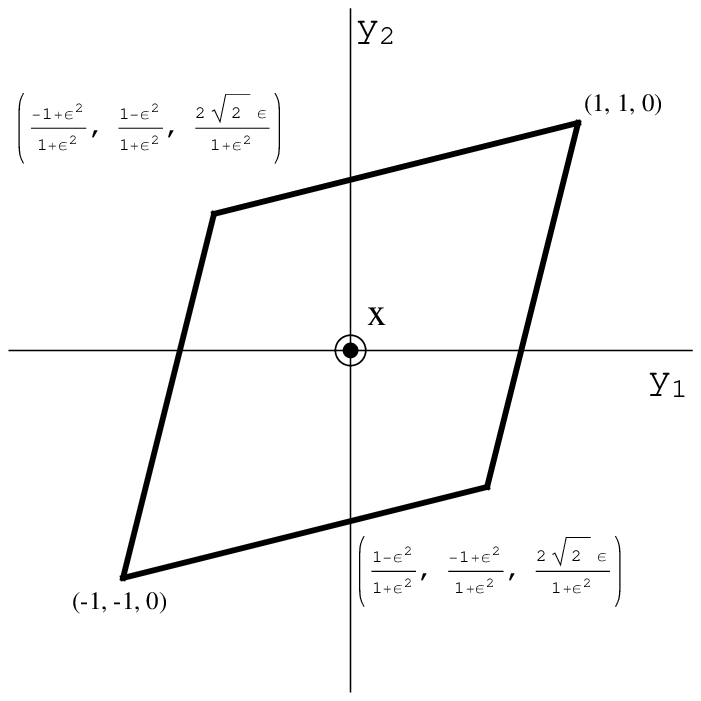}}
\centerline{\ninepoint\sl \baselineskip=2pt {\bf Figure 5:}
{\sl The bounding time-like line segments of the four-cusp $qqqq$ }}
\centerline{\ninepoint\sl all-masses-equal worldsheet viewed in the $y_1,y_2,x$ subspace.}
\bigskip

So far, all we have done is to construct a polygon in the $y_1/y_2$ plane which could be interpreted as a $qqqq$ worldsheet boundary. This polygon is shown in figure 5. The polygon is made up of the original massive cusp and its mirror image. We already know that the original massive cusp is the image of a light-like cusp at $x=\epsilon$ under the map \AMtransform. What about the mirror image cusp? It turns out that this is also related to a light-like cusp. To see this, note that \imagebdy\ can be obtained from \masscuspbdy\ by taking $\epsilon\to{1\over\epsilon}$. This means that \imagebdy\ is also a solution to \poincarebdyii, but with $\epsilon\to{1\over\epsilon}$. So the mirror image cusp maps under the (inverse of) transformation \AMtransform\ to a light-like cusp lying at $x={1\over\epsilon}$. The light-like cusps at $x=\epsilon$ and $x={1\over\epsilon}$ intersect at infinity. (The same can be said of any two parallel light-like cusps regardless of their $x$ position.) This intersection is revealed by the transformation \AMtransform, which produces massive cusps which intersect at the points $(y_1,y_2)=(1,1)$ and $(-1,-1)$.

Figure 5 shows the worldsheet boundary in the case where all four quark masses are equal, but it is easy to generalize this to the case where two of the quarks have the same mass while the other two share a different mass. This is achieved by allowing the two light-like cusps to lie at any two values of $x$, which we call $\epsilon_l$ and $\epsilon_u$, and we assume without loss of generality that $\epsilon_u>\epsilon_l$. The four vertices depicted in figure 5 then generalize to

\eqn\noname{\left({-1+\epsilon_l^2\over1+\epsilon_l^2},{1-\epsilon_l^2\over1+\epsilon_l^2},{2\sqrt{2}\epsilon_l\over1+\epsilon_l^2}\right),\quad \left({-1+\epsilon_u^2\over1+\epsilon_u^2},{1-\epsilon_u^2\over1+\epsilon_u^2},{2\sqrt{2}\epsilon_u\over1+\epsilon_u^2}\right),\quad (1,1,0),\quad (-1,-1,0).}
$y_0$ along the boundary can be obtained from \yzero. As before, this boundary corresponds to a special kinematic regime which we can generalize by performing a boost in the $Y0/Y4$ plane with boost parameter $\beta$. Following the analysis of section 3, we find that the momenta are given by

\eqn\noname{\eqalign{2\pi\alpha' k^\mu_1&= {2\over1-\beta^2}\left(\sqrt{1+\beta^2},{\beta-\epsilon_u^2\over1+\epsilon_u^2},{-1+\beta\epsilon_u^2\over1+\epsilon_u^2}\right),\cr 2\pi\alpha' k^\mu_2 &={2\over1-\beta^2}\left(-\sqrt{1+\beta^2},{\epsilon_l^2-\beta\over1+\epsilon_l^2},{1-\beta\epsilon_l^2\over1+\epsilon_l^2}\right),\cr 2\pi\alpha' k^\mu_3 &= {2\over1-\beta^2}\left(\sqrt{1+\beta^2},{1-\beta\epsilon_l^2\over1+\epsilon_l^2},{\epsilon_l^2-\beta\over1+\epsilon_l^2}\right),\cr 2\pi\alpha' k^\mu_4&={2\over1-\beta^2}\left(-\sqrt{1+\beta^2},{-1+\beta\epsilon_u^2\over1+\epsilon_u^2},{\beta-\epsilon_u^2\over1+\epsilon_u^2}\right).}}
The masses associated with these momenta are

\eqn\qqqqmasses{m_1^2=m_4^2={8\epsilon_u^2\over(2\pi\alpha')^2(1-\beta)^2(1+\epsilon_u^2)^2},\qquad m_2^2=m_3^2={8\epsilon_l^2\over(2\pi\alpha')^2(1-\beta)^2(1+\epsilon_l^2)^2},}
and the Mandelstam variables are given by

\eqn\qqqqmandelstam{\eqalign{s&=-(k_1+k_2)^2=-{8(\epsilon_u^2-\epsilon_l^2)^2\over(2\pi\alpha')^2(1-\beta)^2(1+\epsilon_l^2)^2(1+\epsilon_u^2)^2},\cr t&=-(k_1+k_4)^2=-{8\over(2\pi\alpha')^2(1+\beta)^2}.}}
The case of $ggqq$ scattering is reproduced by sending $\epsilon_u\to\infty$. $gggg$ scattering results are obtained by also taking $\epsilon_l\to0$.

These arguments lead us to consider surfaces in the original $AdS$ space (before \AMtransform\ is applied) which end on two light-like cusps located at different values of $x$. This would be a generalization of the Berkovits-Maldacena solution ending on a single cusp which was employed in section 4 to produce the $ggqq$ worldsheet. In this case, solutions of the form $x(y_0,y_1)$ will not be single-valued. In particular, we should expect from the symmetries of the problem that $x(y_0,y_1)$ will be a double-valued function for a worldsheet extending between two cusps at different values of $x$. To avoid dealing with a double-valued function, we can instead solve for a function $T(x)$ where $T=\sqrt{y_0^2-y_1^2}$. Rewriting the Nambu-Goto Lagrangian in terms of $T(x)$, one finds

\eqn\noname{{\cal L}={1\over2\pi\alpha'}{T\over x^2}\sqrt{T'^2-1},}
with equation of motion

\eqn\Teom{{x T T''+(2 T T'-x)(T'^2-1)\over x^3(T'^2-1)^{3/2}}=0.}
We are interested in solutions which end on light-like cusps at two different values of $x$. This means that $T(x)$ has (at least) two roots, one at $x=\epsilon_l$ and one at $x=\epsilon_u$. If the two roots are inverses of each other, then all four quarks will have the same mass. More general root locations should lead to $qqqq$ scattering amplitudes for which two of the quarks have one mass and the other two have a different mass.

Happily and perhaps surprisingly, exact solutions for $T(x)$ exhibiting the desired root structure can be obtained by joining together worldsheet solutions which end on a single cusp. Two different classes of such solutions have already been found, one by Berkovits and Maldacena \BM\ and the other by Sommerfield and Thorn (ST)\ST. $T(x)$ can be constructed by joining a solution of the BM class to a solution of the ST class, and the resulting worldsheet is smooth. Furthermore, the $T(x)$ obtained in this way has exactly two roots which can assume any positive values. A detailed review of these classes of single-cusp solutions can be found in appendix A. In the following, we will highlight only the most important results from the appendix.

As for the single-cusp solution leading to the $ggqq$ scattering worldsheet, all the BM and ST single-cusp solutions can be constructed by working with the hyperbolic coordinates $w$ and $\sigma$ defined in \hyperboliccoords, and by looking for solutions of the form $x=w e^{\tau(w)}$. The full set of solutions can be written in the form

\eqn\tauofw{\tau(w)=-\half\log(w^2-1)+\eta G(w)+\log\epsilon.}
$\eta$ takes the values $\pm1$, where $\eta=+1$ corresponds to ST-type solutions and $\eta=-1$ corresponds to BM-type solutions. $G(w)$ is a complicated expression involving elliptic functions which can be found in the appendix, but the precise form of $G$ will not be necessary for the present discussion. The only facts about $G$ that we will need here are that it is real, finite and strictly negative in the range $(w_0,\infty)$ for some $w_0>\sqrt{2}$, and that $G$ also depends on a continuous parameter which we can choose to be $w_0$. It is also true that $G$ vanishes in the limit $w\to\infty$, and its first derivative diverges at $w=w_0$.

We can construct solutions for $T(x)$ with roots at $\epsilon_l$ and $\epsilon_u$ from \tauofw. This can be seen by considering for example the $x(w)$ which results from \tauofw:

\eqn\xofw{x(w)=\epsilon{w\over\sqrt{w^2-1}}e^{\eta G(w)}.}
Recalling that the light-like cusp boundary is approached by sending $w\to\infty$, we see that the vanishing of $G$ in this limit implies $x=\epsilon$. Therefore, all of the solutions in \xofw\ end on light-like cusps at $x=\epsilon$. Furthermore, a closer look at \xofw\ reveals that solutions with $\eta=-1$ extend upward from the cusp toward larger values of $x$, while solutions with $\eta=+1$ extend downward from the cusp. This suggests the possibility of constructing a worldsheet stretching between light-like cusps at two different values of $x$ by joining a solution with $\eta=+1$ to a solution with $\eta=-1$. Consider the following two solutions:

\eqn\twosolns{x_u=\epsilon_u {w\over\sqrt{w^2-1}}e^{G(w)},\qquad x_l=\epsilon_l {w\over\sqrt{w^2-1}}e^{-G(w)}.}
The first solution ends on a light-like cusp at $x=\epsilon_u$ and extends downward, while the second solution ends on a light-like cusp at $x=\epsilon_l$ and extends upward. So long as the two solutions are chosen to have the same value of $w_0$ with $w_0$ chosen such that $\epsilon_u/\epsilon_l=e^{-2G(w_0)}$, then the two solutions will join smoothly at $w=w_0$,\foot{Smoothness follows from the fact that the two solutions are just different branches of a single solution to a nonlinear second-order differential equation. In other words, choosing $\epsilon_u$ and $\epsilon_l$ in this way fixes the boundary conditions necessary to uniquely specify a smooth $T(x)$ which solves \Teom.} producing a smooth worldsheet stretching between the two cusps. The two solutions join together at $x_{join}=\sqrt{\epsilon_u\epsilon_l}w_0/\sqrt{w_0^2-1}$.

Although the two-cusp solutions we have described yield exact solutions to \Teom, we do not have a closed form for $T(x)$, so we rely on numerics for a study of its properties. $T(x)$ can be constructed numerically by plotting $T=e^{\tau(w)}$ versus $x(w)$ for each of the solutions $x_l(w)$ and $x_u(w)$ over the range $(w_0,\infty)$. Figure 6 shows $T(x)$ for a particular choice of roots, but the qualitative behavior is the same for any choice of $\epsilon_u>\epsilon_l$.

\bigskip
\centerline{\epsfxsize=0.40\hsize\epsfbox{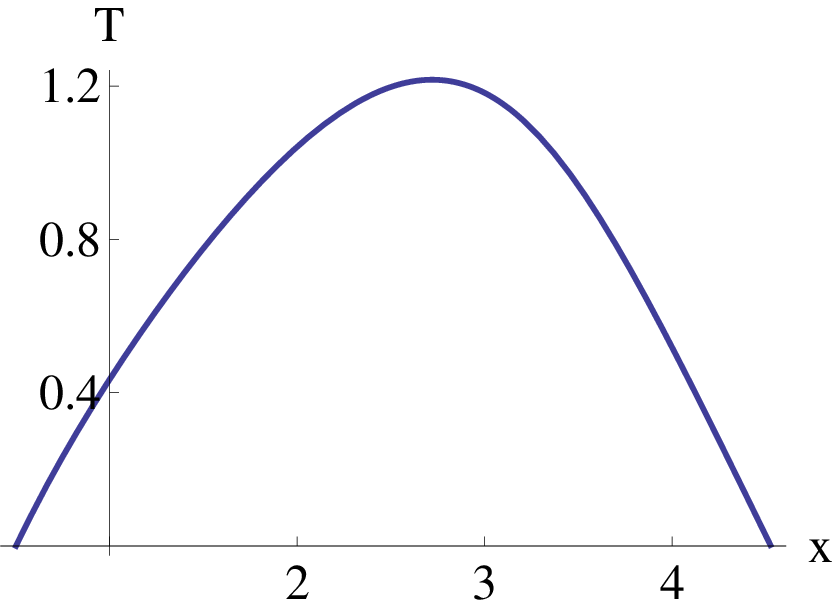}}
\centerline{\ninepoint\sl \baselineskip=2pt {\bf Figure 6:}
{\sl $T(x)$ with $\epsilon_l=0.50$ and $\epsilon_u=4.52$.}}
\bigskip

\noindent For example, it is always the case that $T(x)$ has only one extremum (obviously always a maximum since $T\ge0$) between $\epsilon_l$ and $\epsilon_u$. Therefore, we divide the interval $(\epsilon_l,\epsilon_u)$ into two intervals $(\epsilon_l,x_{max})$ and $(x_{max},\epsilon_u)$, with $x_{max}$ denoting the value of $x$ at which $T(x)$ attains its maximum. We may then invert $T(x)$ on each of these intervals to obtain $x_L(T)$ and $x_U(T)$. Replacing $T$ by $\sqrt{y_0^2-y_1^2}$ in $x_L$ and $x_U$ allows us to plot the two-cusp worldsheet in the $AdS_3$ subspace parametrized by $y_0$, $y_1$ and $x$. This is shown in figure 7. It is important to note that $x_{max}$ is not the same as the joining point of the two solutions, $x_{join}$, so that $x_U$ and $x_L$ are distinct from $x_u$ and $x_l$. Indeed, if we attempted to construct $x_u(T)$, we would run into difficulty because $x_u(T)$ is not single-valued.

\bigskip
\centerline{\epsfxsize=0.30\hsize\epsfbox{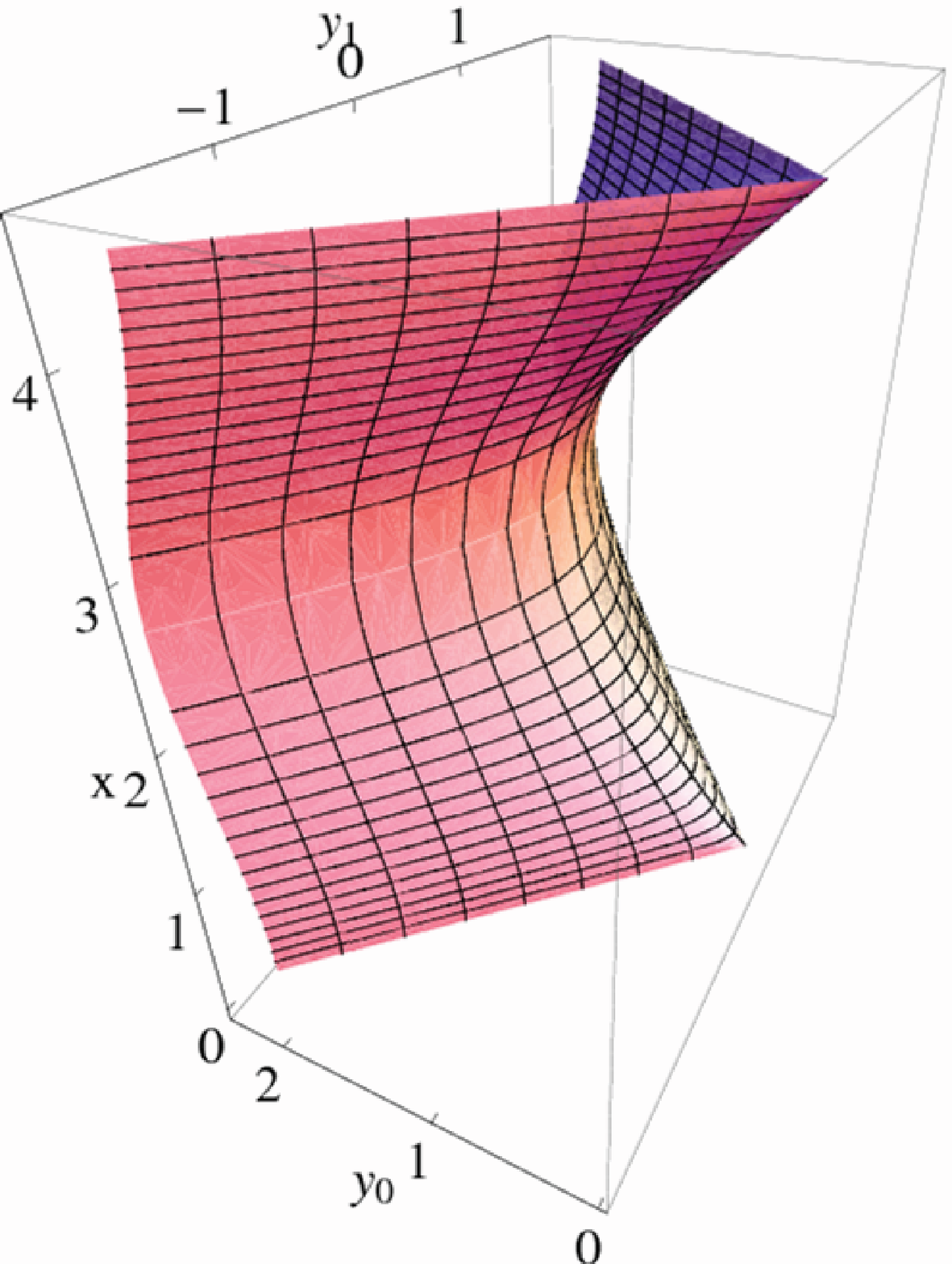}}
\centerline{\ninepoint\sl \baselineskip=2pt {\bf Figure 7:}
{\sl The worldsheet extending between two light-like cusps at $x=1/2$ and $x=4.53$.}}
\bigskip

We are now ready to apply the map \AMtransform\ to obtain a four-cusp worldsheet describing four-quark scattering. By writing $y_0'=\sqrt{T^2+y_1'^2}$ and $x'=x(T)$ in \poincaremapping, we can parametrically generate the transformed two-cusp worldsheet using the solutions we have for $x_L(T)$ and $x_U(T)$. To obtain the entire worldsheet, we need two copies of \poincaremapping, one with $x'=x_L$ and one with $x'=x_U$. Examples of massive four-cusp worldsheets produced in this way are shown in figures 8 and 9.

\bigskip
$$\matrix{\epsfxsize=0.35\hsize\epsfbox{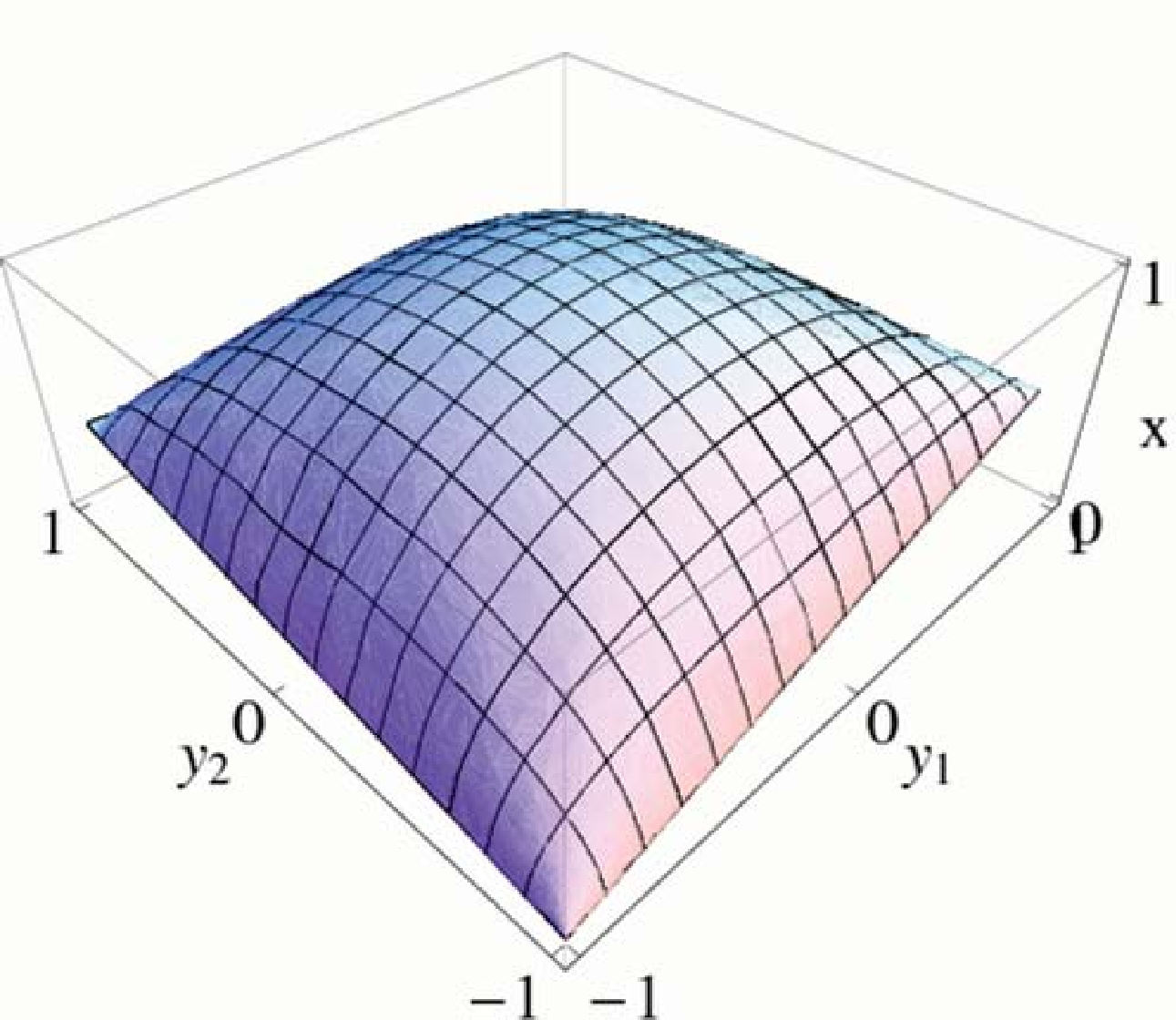} & \qquad & \epsfxsize=0.35\hsize\epsfbox{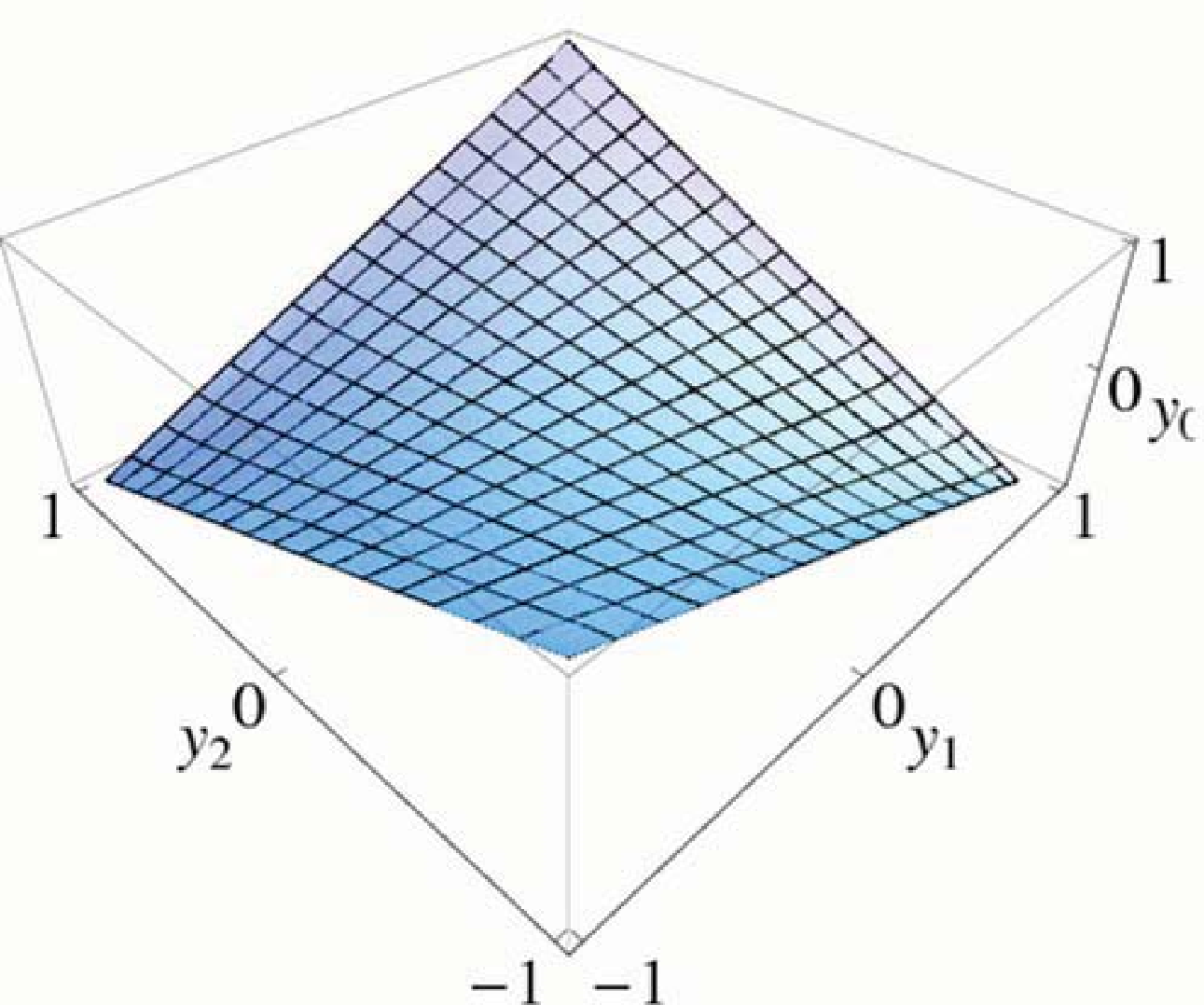}}$$
\centerline{\ninepoint\sl \baselineskip=2pt {\bf Figure 8:}
{\sl $x(y_1,y_2)$ and $y_0(y_1,y_2)$ for $\epsilon_l=0.125$ and $\epsilon_u=5.53$.}}
\bigskip

\bigskip
$$\matrix{\epsfxsize=0.35\hsize\epsfbox{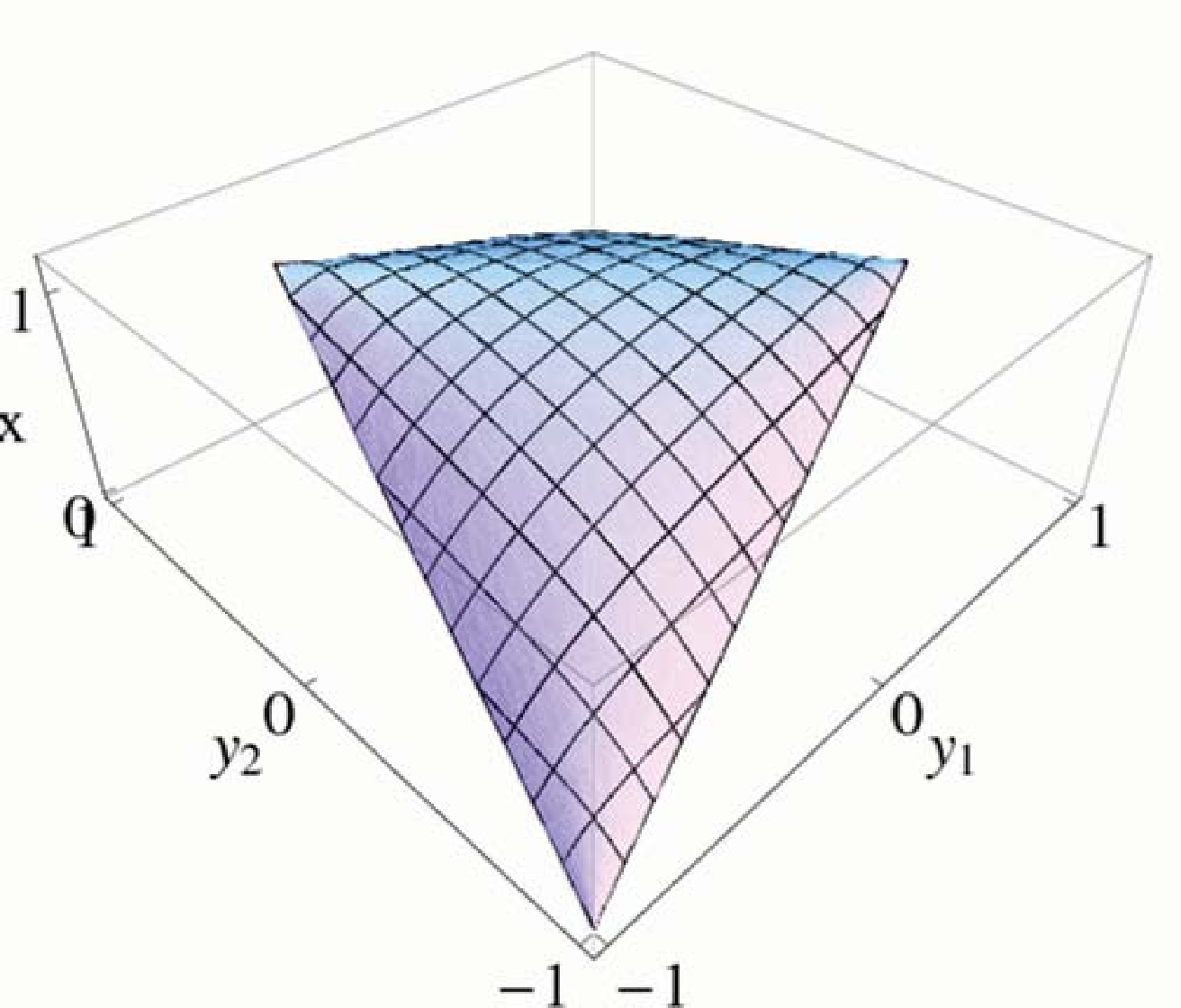} & \qquad & \epsfxsize=0.35\hsize\epsfbox{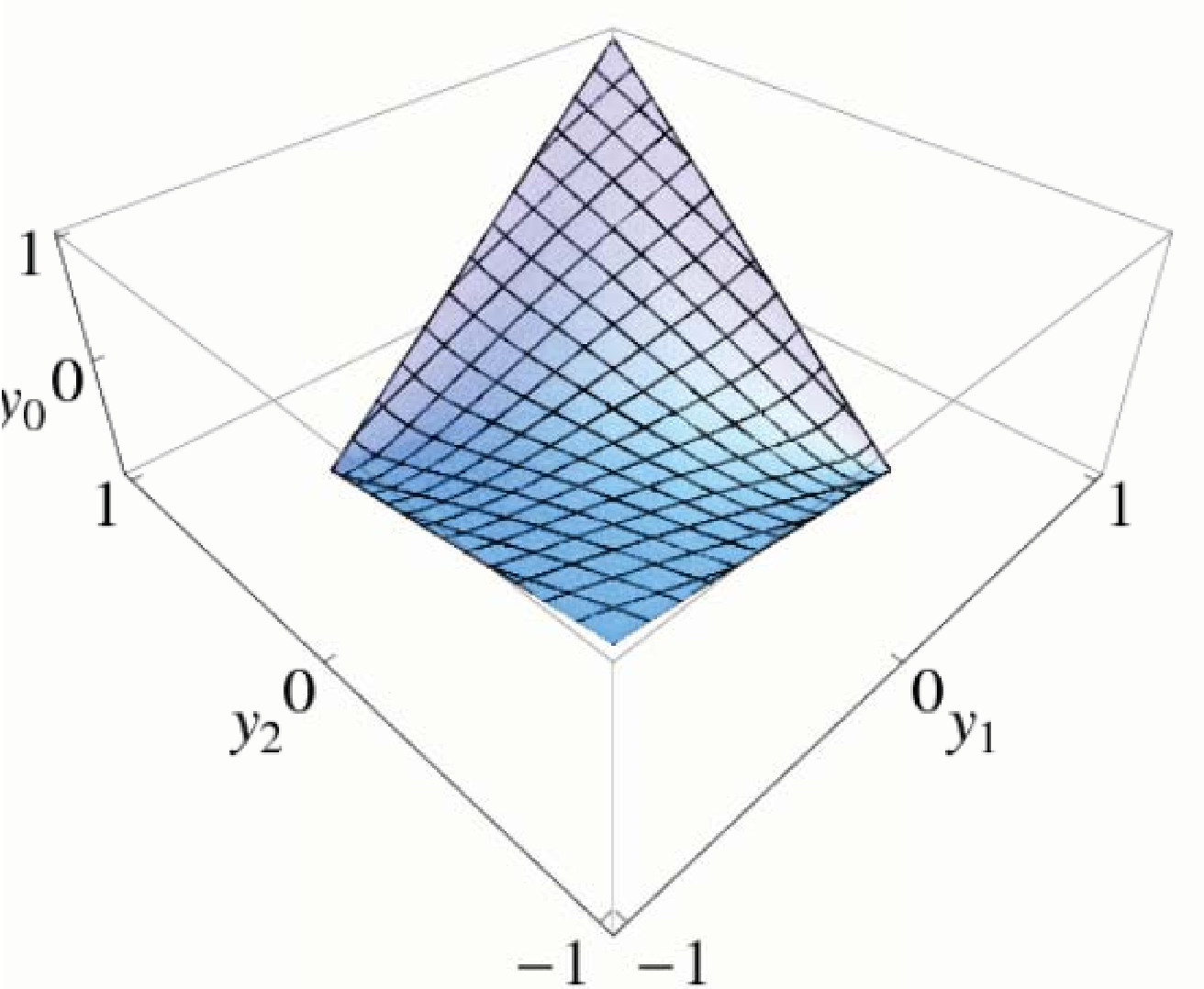}}$$
\centerline{\ninepoint\sl \baselineskip=2pt {\bf Figure 9:}
{\sl $x(y_1,y_2)$ and $y_0(y_1,y_2)$ for $\epsilon_l=0.50$ and $\epsilon_u=1.96$.}}
\bigskip

The Nambu-Goto action evaluated on the solution \xofw\ and regulated with the radial cutoff $x_c$ is

\eqn\qqqqaction{S_{\eta,\epsilon}={iw_0w_*\over\pi\alpha'}\int_{w_0}^\infty dw {\Sigma(w)\over w^2 \sqrt{w^2-w_0^2}\sqrt{w^2-w_*^2}},}
with $w_*\equiv w_0/\sqrt{w_0^2-1}$ and $\Sigma(w)$ is given by

\eqn\noname{\Sigma(w)=\cosh^{-1}\left[{\sqrt{2}w\over(1+\beta)x_c}-{(1-\beta)\sqrt{w^2-1}\over(1+\beta)}\cosh\left(\eta G(w)+\log\epsilon\right)\right].}
The action for the full worldsheet is given by the sum of the contributions from $x_u$ and $x_l$:

\eqn\noname{S_{qqqq}=S_{1,\epsilon_u}+S_{-1,\epsilon_l}.}
Note that unlike the case of $ggqq$ scattering, the integration limits in \qqqqaction\ do not depend on the radial cutoff. This is because the four-quark worldsheet boundary lies almost entirely above $x=x_c$ (only infinitesimal regions near the two vertices at $x=0$ lie below the cutoff), whereas more than half of the $ggqq$ boundary lies below the cutoff. Because $G(w)$ is finite for all $w$ in the range $(w_0,\infty)$ (so long as $w_0>\sqrt{2}$), we can expand in powers of $x_c$\foot{This limit assumes that $x_c/\alpha'$ is much smaller than the quark masses. In the opposite limit, one would expect to get back the 4-gluon amplitude \Sggggii, and we have checked numerically that this is indeed the case.}:

\eqn\noname{\Sigma(w)= -\log x_c+\log\left(2\sqrt{2}w\over(1+\beta)\right)+O(x_c).}

\eqn\noname{\eqalign{S_{\eta,\epsilon}&={iw_0w_*\over\pi\alpha'}\log\left(2\sqrt{2}\over(1+\beta)x_c\right)\int_{w_0}^\infty {dw\over w^2 \sqrt{w^2-w_0^2}\sqrt{w^2-w_*^2}}\cr &+{iw_0w_*\over\pi\alpha'}\int_{w_0}^\infty dw {\log w\over w^2 \sqrt{w^2-w_0^2}\sqrt{w^2-w_*^2}}.}}
In the limit $x_c\to0$, the dependence of the action on $\eta$ and $\epsilon$ drops out, so that the contributions from $x_u$ and $x_l$ are equal, and we no longer need to distinguish between them. Note that the action still depends on the ratio $\epsilon_u/\epsilon_l$ through its dependence on $w_0$ via the relation $\epsilon_u/\epsilon_l=e^{-2G(w_0)}$. Performing the first integration and simplifying slightly the second, the total action becomes

\eqn\noname{\eqalign{S_{qqqq}&={2i\sqrt{w_0^2-1}\over\pi\alpha'w_0}\log\left(2\sqrt{2}w_0\over(1+\beta)x_c\right)\left[K\left(1\over w_0^2-1\right)-E\left(1\over w_0^2-1\right)\right]\cr &+{i\over2\pi\alpha'w_0\sqrt{w_0^2-1}}I\left({w_0^2-2\over w_0^2-1}\right).}}
$K$ and $E$ are complete elliptic integrals of the first and third kinds respectively, and

\eqn\noname{I(a)\equiv \int_0^\infty dz {\log(1+z)\over (1+z)^{3/2} \sqrt{z}\sqrt{z+a}}.}
$I(a)$ is a finite and weakly varying function of $a$. In particular, as $a$ varies from 0 to 1, $I(a)$ varies monotonically from 0.93 to 0.61. However, due to the elliptic functions in $S$ and the fact that we do not have a closed form for $I(a)$, it is difficult to express $S$ as a function of the quark masses and Mandelstam variables. In order to obtain an expression of this form, we will consider the small mass limit. More precisely, we will consider the limits $\epsilon_l\ll 1$ and $\epsilon_u\gg 1$. This limit is different from the case of $ggqq$ scattering because we keep $\epsilon_l/\epsilon_u\gg x_c$. When $\epsilon_l/\epsilon_u$ is small, $w_0\approx\sqrt{2}$. As $w_0$ approaches $\sqrt{2}$, every term in $S$ remains finite except for the elliptic function $K$, which diverges logarithmically. If we expand the function $G(w_0)$ about $w_0=\sqrt{2}$, we find

\eqn\noname{G(w_0)=-\sqrt{2}K\left(1\over w_0^2-1\right)-\log(\sqrt{2}-1)+O(w_0-\sqrt{2}).}
Therefore, using the relation between $G(w_0)$ and $\epsilon_u/\epsilon_l$, we may write

\eqn\noname{K\left(1\over w_0^2-1\right)\approx-{G(w_0)\over\sqrt{2}}-{\log(\sqrt{2}-1)\over\sqrt{2}}={1\over2\sqrt{2}}\log{\epsilon_u\over\epsilon_l}-{\log(\sqrt{2}-1)\over\sqrt{2}}.}
Thus in the limit $\epsilon_l\ll\epsilon_u$, the action is given by

\eqn\noname{S_{qqqq}\approx -{i\over\pi\alpha'}\log\left((1+\beta)x_c\over4\right)\left[{1\over2}\log{\epsilon_u\over\epsilon_l}-\log(\sqrt{2}-1)-\sqrt{2}-1\right]+{i\over2\sqrt{2}\pi\alpha'}I(0).}
A review of \qqqqmasses\ and \qqqqmandelstam\ reveals that we may write

\eqn\noname{(1+\beta)x_c\equiv {4\mu\over\sqrt{-t}},\qquad {\epsilon_u\over\epsilon_l}={-s\over m_1m_3},}
where we have defined the IR cutoff $\mu$ as in section 4, namely $\mu\equiv x_c/(2\sqrt{2}\pi\alpha')$. Using also that $I(0)=\half\pi^2-4$ and $\alpha'=\sqrt{\lambda}$, the action then becomes

\eqn\noname{-iS_{qqqq}\approx{\sqrt{\lambda}\over\pi}\log\left(\mu\over\sqrt{-t}\right)\left[\log\left({\sqrt{ m_1m_3}\over\sqrt{-s}}\right)+\log(\sqrt{2}-1)+\sqrt{2}+1\right]+const.}

The IR divergence appearing in $S_{qqqq}$ was built-in from the beginning of the calculation. It arises from the fact that the Lagrangian evaluated on the $qqqq$ worldsheet solution is independent of the worldsheet coordinate $\sigma$ so that the action contains the overall factor $\int_{-\infty}^\infty d\sigma$. A similar single-log divergence also arises in massive parton scattering in QCD \CataniEF. Note that in the case where $m_1=m_3$, we must include in $S_{qqqq}$ the contribution of an additional worldsheet. This second worldsheet is essentially the same as the one we have constructed above, but with $s$ and $t$ exchanged. The relation between the two worldsheets is analogous to a crossing symmetry in field theory.


\bigskip

\noindent{\bf Acknowledgments}

\noindent We would like to thank Ed Yao and Peter Arnold for helpful discussions.

\appendix{A}{Exact single-cusp world sheet solutions}
A class of worldsheet solutions was found by Berkovits and Maldacena (BM) in appendix B of \BM. These solutions end on two intersecting light-like lines which lie on a plane located a finite distance away from the $AdS_5$ boundary at $x=0$. We can think of these solutions as single-cusp solutions which have been regulated using an IR regulator $D3$-brane which we take to be located at $x=\epsilon$. One of these solutions is used in section 4 to construct $ggqq$ scattering worldsheets. We will in fact consider a simple extension of the set of solutions derived in \BM. The additional solutions we will study were found by Sommerfield and Thorn (ST) \ST. We will show that these solutions also end on light-like cusps located at a finite value of $x$. However, they cannot be thought of as IR regulated versions of worldsheets ending on cusps at $x=0$ because the ST worldsheets hang below the cusps on which they end. By combining BM solutions with ST solutions, it is possible to construct exact worldsheet solutions describing four-quark scattering, which is the focus of section 5 of this paper. In this appendix, we will review the derivation and properties of both sets of solutions.

We begin by choosing coordinates $\tau$ and $\sigma$ such that

\eqn\noname{y_0=e^\tau\cosh(\sigma),\qquad y_1=e^\tau\sinh(\sigma),}
and we suppose that $x$ is only a function of $\tau$ and is taken to be of the form

\eqn\noname{x=w(\tau)e^\tau.}
The Nambu-Goto Lagrangian reduces to

\eqn\wlagrange{{\cal L}\sim {\sqrt{1-(w+w')^2}\over w^2}.}
Since this has no explicit $\tau$ dependence, the corresponding Hamiltonian is conserved:

\eqn\conslaw{H={w(w+w')-1\over w^2\sqrt{1-(w+w')^2}}=const.}
This conservation law allows us to solve for $w'$:

\eqn\wprime{w'={1-w^2-H^2w^4+\eta\sqrt{H^4w^6+H^2w^4-H^2w^2}\over H^2w^3+w},}
where $\eta=\pm1$.

BM actually only considered solutions with $\eta=-1$. Solutions with $\eta=+1$ were analyzed by ST \ST. We will allow for both possible signs in what follows. The case $H^2=-1/4$ was analyzed both by BM\foot{Note that BM had an extra factor of $i$ in the definition of the Lagrangian \wlagrange, which is why their constant of motion $c^2=-H^2$.} and by ST, and it is of particular importance. BM showed that this solution is the only one which asymptotes to the single-cusp solution of AM \AM. This is evidenced by the fact that the AM solution also obeys $H^2=-1/4$. Thus, in the limit where the IR regulator brane rejoins with the stack of $D3$-branes, i.e. when $\epsilon\to0$, the BM solution with $H^2=-1/4$ becomes the AM single-cusp solution.

For $H^2=-1/4$, \wprime\ becomes

\eqn\noname{w'=-{(w^2-2)(w-\eta)\over w(w-2\eta)}.}
This is easily integrated to yield an implicit form for $w(\tau)$:

\eqn\tauBM{Ce^\tau={1\over w-\eta}\left(w-\sqrt{2}\over w+\sqrt{2}\right)^{{\eta\over\sqrt{2}}}.}
Setting $\eta=-1$ gives the IR brane-regulated solution of BM, for which we take $C=1/\epsilon$. It is this solution that we employ in section 4 to obtain the $ggqq$ worldsheet. Equation \BMsinglecusp\ is obtained from \tauBM\ by rewriting $w$ and $\tau$ in terms of $x$ and $\sqrt{y_0^2-y_1^2}$ and re-arranging the result.

The solution with $\eta=+1$ also ends on a cusp at $x=\epsilon$. However, unlike the solution studied by BM, this solution extends downward from the cusp toward smaller values of $x$. In fact, this solution ends on a second light-like cusp at $x=0$, as is evident from the fact that $\tau\to-\infty$ at both $w=\infty$ and $w=\sqrt{2}$. This solution is also related to a $ggqq$ worldsheet which is essentially the same worldsheet as that obtained from the BM solution in section 4, but with the quark pair swapped with the gluon pair.

ST focused on the $\eta=+1$ solution for which $1<w\le\sqrt{2}$, in which case one must take the constant $C=(-1)^{1/\sqrt{2}}\times real$. This solution does cover the interior of the cusp. However, it does not end on a cusp lying at a finite value of $x$ but instead ends on a light-like cusp at $x=0$, so this solution does not play a role in the study of massive particle scattering.\foot{We should also point out that, unlike the AM single-cusp solution, this solution does not lead to a four-gluon solution owing to the behavior of the solution at infinity. Details can be found in \ST.}

Since the AM solution is thought to be unique, we expect that the remainder of the solutions with $H^2\ne-1/4$ will not admit an interpretation as IR-regulated worldsheets ending on a light-like cusp at the $AdS$ boundary. We will now show that this is indeed the case. We would first like to determine how many of these solutions correspond to a worldsheet which ends on a light-like cusp lying at $x=\epsilon>0$. This cusp is located at $\tau=-\infty$. Since $x=we^\tau$, it must be that $w\sim e^{-\tau}$ as $\tau\to-\infty$ if $x$ is to go to a positive constant in this limit. Taking $w\to\infty$ in \wprime\ reveals that $w'\to-w$, so that indeed $w=\epsilon e^{-\tau}$ irrespective of the value of $H^2$. Therefore, we conclude that any value of $H^2$ will yield a solution ending on a light-like cusp lying at $x=\epsilon$.

However, requiring the worldsheet to end on a light-like cusp at $x=\epsilon$ is not sufficient to ensure that the worldsheet will have the appropriate boundary behavior as $\epsilon\to0$. That is, we still need to check that the solutions extend all the way to $\tau=\infty$. In particular, it is possible that some of these solutions will end at a finite value of $\tau$, resulting in an extra unwanted worldsheet boundary. Such solutions exhibiting an extra boundary should not be counted along with the AM solution in a computation of the cusp amplitude. In fact, it turns out that only the $H^2=-1/4$ solution found by BM extends all the way to $\tau=\infty$.\foot{At this point, we might also consider eliminating worldsheets that extend downward instead of upward from the $x=\epsilon$ plane, but this turns out to not be necessary as the condition that the worldsheet extends all the way to $\tau=\infty$ is strong enough to eliminate all but one solution.} The simplest way to see this is to integrate \wprime\ directly. The result is

\eqn\taui{\tau(w)=-{1\over2}\log(w^2-1)+\eta G(w)+\log\epsilon,}
where we have defined

\eqn\Gofw{\eqalign{G(w)&\equiv {i\over w_0^2-1}\bigg\{w_0 F\left(i\sinh^{-1}\left(w_0\over\sqrt{w^2-w_0^2}\right),1-{w_*^2\over w_0^2}\right)\cr &-{1\over w_0}\Pi\left({w_0^2-1\over w_0^2},i\sinh^{-1}\left({w_0\over \sqrt{w^2-w_0^2}}\right),1-{w_*^2\over w_0^2}\right)\bigg\},}}
and

\eqn\noname{w_0^2\equiv {-1+\hbox{sgn}(H^2)\sqrt{1+4H^2}\over2H^2},\qquad w_*^2\equiv {-1-\hbox{sgn}(H^2)\sqrt{1+4H^2}\over2H^2}.}
Notice that $w_0^2\ge w_*^2$ for all $H^2\ge-1/4$. $F$ is an incomplete elliptic integral of the first kind, while $\Pi$ is an incomplete elliptic integral of the third kind. The integration constant was chosen to be $\log\epsilon$ so as to agree with the behavior $w\to\epsilon e^{-\tau}$ in the limit $\tau\to-\infty$, $w\to\infty$. It is straight-forward to check that the solution is real for $w\ge w_0$. In order for the solution to completely cover the interior of the cusp, it would have to be the case that $\tau\to\infty$ for some value of $w\ge w_0$. However, \taui\ is finite for $w\ge w_0$ unless $H^2=-1/4$ or $H=0$, in which cases \taui\ diverges at $w=w_0$. Figure 10 shows typical behavior of $\tau(w)$ for each choice of $\eta=\pm1$. The solution with $H=0$ will be studied further in appendix B, where it will be shown that the worldsheet is time-like so that the associated scattering amplitude receives a phase instead of an exponential suppression factor. Thus, only the solution with $H^2=-1/4$ can properly be thought of as a regulated version of a worldsheet ending on a light-like cusp at $x=0$, in support of the conjectured uniqueness of the AM solution.

\bigskip
$$\matrix{\epsfxsize=0.35\hsize\epsfbox{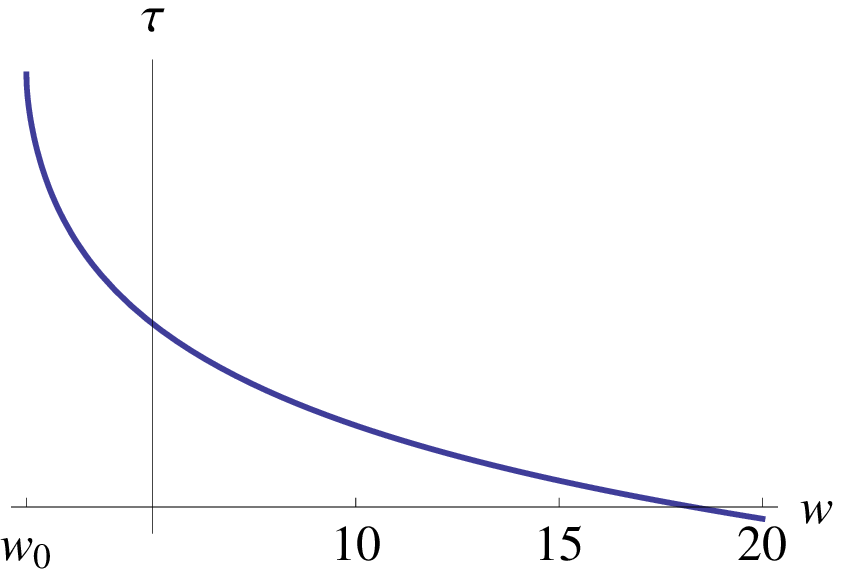} & \qquad & \epsfxsize=0.35\hsize\epsfbox{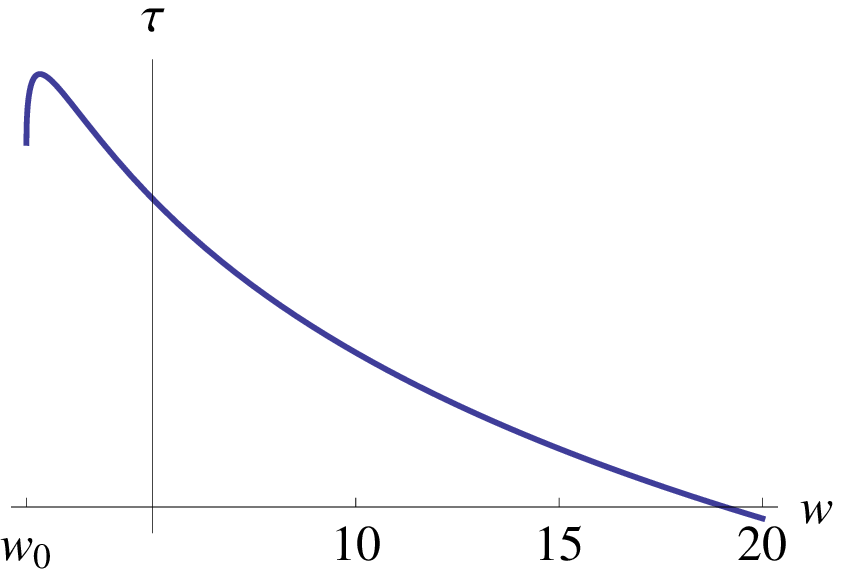}}$$
\centerline{\ninepoint\sl \baselineskip=2pt {\bf Figure 10:}
{\sl $\tau(w)$ for $\eta=-1$ (left) and $\eta=+1$ (right) with $H^2=-1/5$.}}
\bigskip

\bigskip
\centerline{\epsfxsize=0.40\hsize\epsfbox{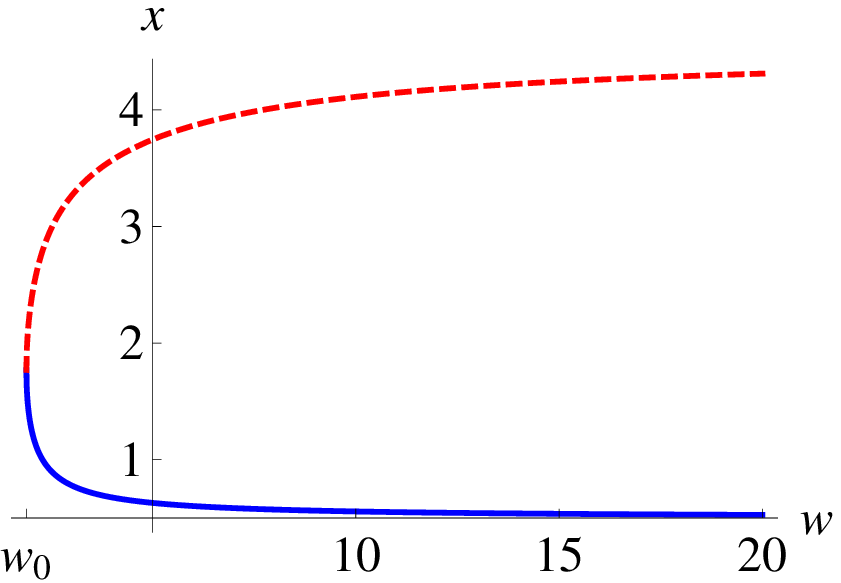}}
\centerline{\ninepoint\sl \baselineskip=2pt {\bf Figure 11:}
{\sl $x(w)$ for $H^2=-1/5$ with $\eta=+1$, $\epsilon=4.53$ (red/dashed)}}
\centerline{\ninepoint\sl and with $\eta=-1$, $\epsilon=1/2$ (blue/solid).}
\bigskip

However, the remaining solutions with $H^2>-1/4$ do have a physical interpretation as revealed in section 5, so we will now examine their properties more closely. We have already noted that the solutions are real in the region $w\ge w_0$. As $w\to\infty$, the worldsheet approaches the light-like cusp at $x=\epsilon$, while at $w=w_0$, the solution appears to abruptly end at a finite value of $\tau$ which we will call $\tau_0$. It is clear from \wprime\ that $w'=0$ at $\tau=\tau_0$. This in turn implies that $|\tau'(w_0)|=\infty$. Note that this is true regardless of the choice of $\eta=\pm1$ in \wprime\ and \taui. That is, both the BM and ST solution sets exhibit this behavior. This is depicted in figure 10.

Figure 11 shows typical behavior for $x(w)$, which can be constructed from $\tau(w)$ using that $x=we^\tau$. Recalling that the light-like cusp is approached as $w\to\infty$, the figure shows the general property that solutions with $\eta=-1$ extend away from the $AdS$ boundary ($x(w)$ decreases as $w$ increases), while solutions with $\eta=+1$ extend toward the $AdS$ boundary ($x(w)$ increases as $w$ increases). Put another way, solutions with $\eta=-1$ extend upward form the cusp toward larger values of $x$, while solutions with $\eta=+1$ hang down from the cusp, extending to smaller values of $x$.

The fact that the two types of solutions, those with $\eta=+1$ and those with $\eta=-1$, extend away from the cusp in different directions suggests the interesting possibility of constructing a solution which ends on two different light-like cusps lying at different values of $x$. This solution would be comprised of a pair of solutions, one with $\eta=+1$ and one with $\eta=-1$. As is clear from figure 11, the total solution will be a connected surface so long as the two solutions have the same value of $H^2$ (and thus the same value of $w_0$) and so long as the parameter $\epsilon$ for each solution is chosen appropriately. In figure 11, the $\epsilon$'s were chosen such that the two solutions have the same value of $x$ at $w_0$. Not only is the resulting total surface connected, it is also smooth, as is guaranteed by the fact that the two solutions are just different branches of a single solution to the nonlinear differential equation \conslaw. An example of such a surface extending between two light-like cusps is shown in figure 7.

A natural question to ask is whether it is possible to find a solution joining two light-like cusps located at any two values of $x$. This can be answered by examining the condition

\eqn\joincondition{\tau(w_0,\eta=1,\epsilon=\epsilon_u)=\tau(w_0,\eta=-1,\epsilon=\epsilon_l).}
Using \taui, the condition \joincondition\ can be restated as

\eqn\noname{\epsilon_u=e^{-2G(w_0)}\epsilon_l.}
We have checked that $G(w_0)$ is a negative, monotonic function of $H^2$ such that as $H^2\to-1/4$, $G\to-\infty$ and as $H\to0_-$, $G\to0$. This then implies that for any choice of $\epsilon_l$, one can obtain any value of $\epsilon_u>\epsilon_l$ by making an appropriate choice of $H^2$ from the range $(-1/4,0)$. We will show below that solutions with $H^2\ge0$ give worldsheets of Minkowskian signature, while solutions with $H^2<0$ give worldsheets with Euclidean signature. Thus, all the solutions with $-1/4<H^2<0$ lead to Euclidean worldsheets extending between two light-like cusps located at any two values of $x$.

In section 5, we use these two-cusp solutions to construct massive four-cusp worldsheets. As explained in section 5, the two-cusp solution can be described by a function $T(x)$ where $T=\sqrt{y_0^2-y_1^2}$. $T(x)$ is closely related to $\tau(w)$ since $T=e^\tau$. It is easy to construct numerically the $T(x)$ associated with the $\tau(w)$ given in \taui\ by making a plot of $T(w)=e^{\tau(w)}$ versus $x(w)=we^{\tau(w)}$. This is the method employed in the construction of the plots in section 5.

For the sake of completeness, we will conclude this section by computing the action for all values of $H^2\ge-1/4$. This is facilitated by rewriting the action as an integral over $w$ rather than $\tau$. The action becomes

\eqn\waction{S= -{1\over2\pi\alpha'}\int d\sigma\int_{w_0}^\infty {dw\over w^2\sqrt{-1+w^2+H^2w^4}}.}
It is helpful to do a coordinate transformation to $u=w^2$:

\eqn\noname{S= -{1\over4\pi\alpha'\sqrt{H^2}}\int d\sigma\int_{u_0}^\infty {du\over u^{3/2}\sqrt{u-u_0}\sqrt{u-u_*}},}
where $u_0\equiv w_0^2$ and $u_*\equiv w_*^2$. The $\sigma$ integration gives rise to a divergence because we have not yet introduced a UV cutoff. Since the four momenta correspond to the legs of the cusp, we will impose the cutoff in the coordinate system whose axes coincide with the cusp:

\eqn\noname{y_\pm=y_0\pm y_1\le L \qquad\Rightarrow\qquad \tau\le\log L, \quad \tau-\log L\le \sigma\le \log L-\tau.}
We then have

\eqn\regaction{S=-{1\over2\pi\alpha'\sqrt{H^2}}\left[\log{L\over\epsilon}\int_{u_L}^\infty {du\over u^{3/2}\sqrt{u-u_0}\sqrt{u-u_*}} -\int_{u_L}^\infty du{\tau(\sqrt{u})-\log\epsilon\over u^{3/2}\sqrt{u-u_0}\sqrt{u-u_*}}\right],}
where $u_L$ is defined such that $u_L\to u_0$ as $L\to\infty$.

Consider first the case $H^2>-1/4$, for which we have that $u_0>u_*$ and $\tau$ is finite for all $u\ge u_0$, so that both integrals in \regaction\ are finite even when we let $u_L\to u_0$. The second integral is then a finite constant independent of $L$ and $\epsilon$ so that we may ignore it compared to the first integral. This first integral evaluates to

\eqn\noname{S\approx {\log{L\over\epsilon}\over \pi\alpha'\sqrt{H^2}\sqrt{u_0}u_*}\left[E\left(u_*\over u_0\right)-K\left(u_*\over u_0\right)\right].}
$K$ is a complete elliptic integral of the first kind, whereas $E$ is a complete elliptic integral of the second kind. $S(H^2)$ is a smooth function of $H^2$ for $-1/4< H^2< 0$. It diverges as $H^2\to-1/4$ and goes to zero as $H^2\to0$ from below. Also note that it is purely imaginary in this region, i.e. the worldsheets for $-1/4\le H^2<0$ are Euclidean. For $H^2\ge0$, $S$ is purely real, diverges as $H^2\to\infty$, and goes to a finite constant as $H^2\to0$ from above so that $S(H^2)$ is discontinuous at $H^2=0$. Therefore, the worldsheets for $H^2\ge0$ are Minkowskian and will only lead to a phase factor for associated scattering amplitudes instead of an exponential suppression. This is why we have largely ignored the $H^2\ge0$ worldsheets in this paper.

We now return to the case $H^2=-1/4$. This time, neither integral in \regaction\ is finite if we allow $u_L\to u_0$, so we must keep $u_L>u_0$. It is possible to perform both integrals when $H^2=-1/4$ since $\tau(w)$ simplifies:

\eqn\noname{\tau(w)=-\log(w-\eta)+{\eta\over\sqrt{2}}\log(w-\sqrt{2})-{\eta\over\sqrt{2}}\log(w+\sqrt{2})+\log\epsilon.}
This follows immediately from \tauBM. It can also be obtained as a limiting case from \taui. Setting $\tau=\log L$ and taking $L/\epsilon$ to be sufficiently large (so that $w\approx\sqrt{2}$), we may write

\eqn\noname{w_L\equiv\sqrt{u_L}=\sqrt{2}+\left(L\over\epsilon\right)^{-\sqrt{2}}.}
After substituting $u_0=u_*=2$, we find for the first integral in \regaction\

\eqn\noname{-{1\over2\pi\alpha'\sqrt{H^2}}\log{L\over\epsilon}\int_{u_L}^\infty {du\over u^{3/2}(u-2)}={i\over2\pi\alpha'}\log^2{L\over\epsilon}+i{{3\over4}\log 2-1\over\sqrt{2}\pi\alpha'}\log{L\over\epsilon},}

\noindent while the second integral evaluates to

\eqn\noname{{1\over2\pi\alpha'\sqrt{H^2}}\int_{u_L}^\infty du{\tau-\log\epsilon\over u^{3/2}(u-2)}={i\eta\over4\pi\alpha'}\log^2{L\over\epsilon}+i\eta{{3\over4}\log2-{1\over2}\sinh^{-1}(1)\over\sqrt{2}\pi\alpha'}\log{L\over\epsilon}+const.}
Adding the values for these two integrals gives to leading order

\eqn\regactionresult{S\approx {i(2+\eta)\over4\pi\alpha'}\log^2{L\over\epsilon}.}
When $\eta=-1$, the worldsheet can be thought of as a regularized version of the single-cusp AM solution, and indeed AM obtained the same value of the action using other regularization schemes. Although the integrals which arise for this regularization scheme are fairly simple, the momentum cutoff $L$ we have employed here does not have a clear interpretation in terms of the four-cusp solution, which is why we favor the radial cutoff scheme in sections 4 and 5.

\appendix{B}{$H=0$ and a time-like massive cusp solution}
In this appendix, we will take a closer look at the $H=0$ solution, which is the simplest solution of the class described in appendix A. The worldsheet has the form

\eqn\Hiszero{x=\sqrt{\epsilon^2+y_0^2-y_1^2}.}
Unlike the $H^2=-1/4$ solution, however, the $H=0$ solution does not lead to an exponential suppression factor for the associated scattering amplitude since the worldsheet is time-like instead of space-like, as is readily seen by plugging \Hiszero\ into the Lagrangian:

\eqn\noname{{\cal L}\sim{\epsilon\over x^3}.}
One could obtain a space-like worldsheet by taking $\epsilon$ to be purely imaginary, but then \Hiszero\ would not correspond to a worldsheet ending on a light-like cusp. It would be interesting to find a physical interpretation for the solution with $\epsilon^2<0$. We should also point out that when $\epsilon=0$, \Hiszero\ is no longer a solution.

Despite the fact that \Hiszero\ does not seem to have much to say about scattering amplitudes, it does have the nice property that, unlike the rest of the solutions parametrized by $H^2$, one can write down an explicit form for $x(y_0,y_1)$. This makes it a nice toy example for testing ideas on how to obtain new solutions from old ones. For example, one can obtain a worldsheet ending on a massive cusp from \Hiszero. To see this, first note that \Hiszero\ can be generalized to

\eqn\Hiszerogen{x=\sqrt{a+by_0+cy_1+y_0^2-y_1^2}}
by performing translations in the $y_0$ and $y_1$ directions. Now, it is possible to choose $a$, $b$, and $c$ such that \Hiszerogen\ obeys the boundary condition \bcs\ corresponding to a massive cusp as follows. Before showing this, we will rewrite \bcs\ in a more accessible way.

First, we choose the point of the cusp to be located at $y_\mu=0$ and $x=x_{D7}$. This implies that we should choose $y_0^\mu$ such that \bcs\ becomes

\eqn\bcsmod{y_\mu={k_\mu\over m}(x-x_{D7}).}
As usual, we set $k_2=k_3=0$ so that $y_2=y_3=0$. In the $y_0/y_1$ plane, the boundary of the massive cusp is comprised of two time-like lines which intersect at the origin. These lines can be obtained by contracting both sides of \bcsmod\ with a vector $\chi_\mu$ which is orthogonal to $k_\mu$:

\eqn\noname{\chi^\mu y_\mu=0.}
Up to an overall rescaling, the $\chi_\mu$ associated with the momentum $k_\mu=(k_0,k_1)$ has components $\chi_\mu=(k_1,k_0)$. We can then solve for $y_0$ as a function of $y_1$ on the boundary:

\eqn\masscuspbdy{y_0=\alpha |y_1|,}
where we define $\alpha\equiv |k_0/k_1|>1$. We have oriented the cusp so that it is symmetric under $y_1\to-y_1$, i.e. the momenta associated with the two lines are $k_\mu^\pm=(k_0,\pm|k_1|)$, and we focus on the cusp in the region $y_0\ge 0$. Inverting \bcsmod\ and replacing $y_1$ in favor of $y_0$ using \masscuspbdy, we obtain

\eqn\masscuspbc{x\Big|_{bdy}=x_{D7}-{\sqrt{\alpha^2-1}\over\alpha}y_0.}
This form of the boundary condition is useful because it is valid on both lines of the boundary.

We are now ready to construct the promised massive cusp. Returning to \Hiszerogen\ and choosing

\eqn\masscuspparams{a=x_{D7}^2,\qquad b=-2x_{D7}{\sqrt{\alpha^2-1}\over\alpha},\qquad c=0,}
we have

\eqn\masscuspsoln{x=\sqrt{x_{D7}^2-2x_{D7}{\sqrt{\alpha^2-1}\over\alpha}y_0+y_0^2-y_1^2}.}
It is easy to check that this satisfies the boundary condition \masscuspbc. Unfortunately, this massive cusp worldsheet inherits the time-like signature of the $H=0$ light-like cusp solution \Hiszero. Indeed, what we have just seen is that a massive cusp worldsheet can be obtained by appropriately slicing \Hiszero.

\appendix{C}{Open string scattering in flat space for strings with mixed boundary conditions}

In this appendix, we generalize computations of string scattering amplitudes in flat space \GrossKZA\GrossGE\ to the case of open strings of finite extent. The string worldsheet corresponding to the appropriate insertion of vertex operators given in terms of embedding coordinates is
\eqn\noname{\eqalign{X^\mu &= -i \sum_{r=1}^N {p^\mu_r\over \pi} \ln|z-\lambda_r|, \cr
Y^m &=  \sum_{r=0}^N {y_r^m\over\pi} \bigg({\rm arctan}{\lambda_{r+1}-x\over y}-{\rm arctan} {\lambda_r-x\over y}\bigg),}}
where $z=x+iy$ is a complex coordinate on the upper half plane.
For concreteness we have considered a string which obeys Neumann boundary conditions in the $X^\mu$ direction and Dirichlet boundary conditions in the $Y^m$ directions: $\partial_{y} X^\mu(z=\bar z)=0$ and  $Y^m(z=\bar z)=y^m$ for $x\in(\lambda_{r+1},\lambda_r)$.
In particular, for the 2-2 scattering, we fix the residual conformal symmetry in the usual manner by choosing $\lambda_{1}=0, \lambda_2=\lambda, \lambda_3=1,\lambda_4=\infty$, and with $\lambda_0 = -\infty$. The worldsheet action in the conformal gauge is a function of $\lambda$
\eqn\noname{
S=\int d^2 z(\partial_z X^\mu\partial_{\bar z} X_\mu+\partial_z Y^m
\partial_{\bar z} Y^m),
}
which can be simplified by using
\eqn\noname{
{\rm arctan}{x\over y}={i\over 2}\log\left({-\bar z\over z}\right) \Rightarrow \partial_z\partial_{\bar z}{\rm arctan}{ x\over y}=0
}
and
\eqn\noname{
\partial_z\partial_{\bar z}\ln(z \bar z)=2\pi\delta^2(z).
}
After integration by parts the string action becomes
\eqn\os{\eqalign{
S&=-\int_{y\ge 0} d^2 z \,\bigg(X^\mu (\partial_x^2 + \partial_y^2) X_\mu + Y^m(\partial_x^2 + \partial_y^2) Y_m\bigg)  +\int_{-\infty}^\infty dx \,Y^m \partial_y Y_m\bigg|_{z=\bar z}
\cr
&=-\sum_{r=1}^4 X^\mu(z=u_r) p_{r \,\mu} + \int_{-\infty}^\infty dx \,Y^m \partial_y Y_m\bigg|_{z=\bar z}
\cr
&=(2 p_1\cdot p_2\ln\lambda+2 p_2\cdot p_3 \ln(1-\lambda)+2(p_1+p_2+p_3)\cdot p_4 \ln R + \sum_{r=1}^4 p_r^2 \ln\epsilon )
\cr
&+2(y_2-y_1)\cdot (y_1-y_0)\ln\lambda + 2(y_3-y_2)\cdot (y_2-y_1) \ln(1-\lambda) - (y_3-y_0)\cdot (y_3-y_0)\ln \tilde R
\cr
&+ ( (y_1-y_0)^2 + (y_2-y_1)^2+(y_3-y_2)^2)\ln\tilde \epsilon,}}
where momentum conservation yields $\sum_{r=1}^4 p_r^\mu=0$, and the on-shell condition requires $p_r^2 + (y_r-y_{r-1})^2=0$. The action N and D terms have been regularized by introducing the $R,\epsilon$, $\tilde R,  \tilde\epsilon$ parameters, with $R,\tilde R\gg1$ and $\epsilon,\tilde\epsilon\ll1$. If the string had only one type of boundary condition, these divergent terms would cancel due to the on-shell condition. We will ignore these divergent terms in what follows (in principle, one could add boundary terms to the original Lagrangian and thus obtain an action which is finite).

The saddle point approximation for the scattering amplitude is given by the extremum of $\exp(-S(\lambda))$. Extremizing \os we find
\eqn\noname{
\lambda = {{\hat p_2\cdot \hat p_1}\over {\hat p_2\cdot \hat p_1+\hat p_3\cdot \hat p_2}}
,}
where we have defined
\eqn\noname{\hat p_r = (p_r^\mu , y_{r+1}^m-y_r^m).}
For the scattering amplitude, the saddle point approximation validity requires that $\hat p_1\cdot \hat p_2$ and $\hat p_2\cdot \hat p_3$ both be large.
Substituting back into the action, we find that the saddle point approximation gives the following value for the scattering amplitude
\eqn\noname{A\sim \exp(-2\hat p_1\cdot\hat p_2 \ln(\hat p_1\cdot\hat p_2)-2\hat p_2\cdot \hat p_3\ln(\hat p_2\cdot \hat p_3)+2(\hat p_1\cdot \hat p_2+\hat p_2\cdot \hat p_3)\ln( \hat p_1\cdot \hat p_2+\hat p_2\cdot \hat p_3)) }
We have obtained a natural generalization of the amplitudes computed in \GrossKZA\GrossGE\ for strings with Neumann boundary conditions. Our result is expressed in terms of ``higher-dimensional'' momenta. If the string obeys Dirichlet boundary conditions in one dimension, and has finite extent, the ``extra-dimension'' momentum component accounts for mass terms in the scattering amplitude.

\listrefs
\end